\documentclass[aps,twocolumn,epsf,floats,pre]{revtex4-1}
\usepackage{graphicx} 
\usepackage{amsmath}
\usepackage{calrsfs}
\usepackage{color}

\newcommand \be {\begin{equation}}
\newcommand \ee {\end{equation}}
\newcommand \bea {\begin{eqnarray}}
\newcommand \eea {\end{eqnarray}}

\begin{document} 

 

%
%
%
%
%

\title{Flocking and turning: a new model for self-organized collective motion}

\author{Andrea Cavagna$^{a,b,c}$}
\author{Lorenzo Del Castello$^{a,b}$}
\author{Irene Giardina$^{a,b,c}$}
\author{Tomas Grigera$^{d,e}$}
\author{Asja Jelic$^{a,b}$}
\author{Stefania Melillo$^{a,b}$}
\author{Thierry Mora$^f$}
\author{Leonardo Parisi$^{a,b}$}
\author{Edmondo Silvestri$^{a,b,g}$}
\author{Massimiliano Viale$^{a,b}$}
\author{Aleksandra M. Walczak$^h$}

\affiliation{
$^a$Istituto Sistemi Complessi (ISC--CNR), Via dei Taurini 19, 00185 Roma, Italy\\
$^b$Dipartimento di Fisica, ``Sapienza'' Universit\'a di Roma,  P.le Aldo Moro 2, 00185 Roma, Italy\\
$^c$Initiative for the Theoretical Sciences, The Graduate Center, 365 Fifth Avenue, New York, NY 10016 USA\\
$^d$ Instituto de Investigaciones Fisicoqu{\'\i}micas Te{\'o}ricas y Aplicadas (INIFTA) and Departamento de F{\'\i}sica, Facultad de Ciencias Exactas, Universidad Nacional de La Plata, c.c. 16, suc. 4, 1900 La Plata, Argentina\\
$^e$CONICET La Plata, Consejo Nacional de Investigaciones Cient{\'\i}ficas y T{\'e}cnicas, Argentina\\
$^f$Laboratoire de Physique Statistique de l'{\'{E}cole} Normale Sup\'erieure, CNRS and Universites Paris VI and Paris VII, 24 rue Lhomond, 75231 Paris Cedex 05, France\\
$^g$Dipartimento di Matematica e Fisica - Universit\'a Roma Tre, Via della Vasca Navale, 84, 00146 Roma, Italy, and\\
$^h$Laboratoire de Physique Th\'eorique de l'\'Ecole Normale Sup\'erieure, CNRS and University Paris VI, 24 rue Lhomond, 75231 Paris Cedex 05, France}

\date{\today} 

\begin{abstract}
Birds in a flock move in a correlated way, resulting in large polarization of velocities. A good understanding of this collective behavior exists for linear motion of the flock. Yet observing actual birds, the center of mass of the group often turns giving rise to more complicated dynamics, still keeping strong polarization of the flock. Here we propose novel dynamical equations for the collective motion of polarized animal groups that account for correlated turning including solely social forces. We exploit rotational symmetries and conservation laws of the problem to formulate a theory in terms of generalized coordinates of motion for the velocity directions akin to a Hamiltonian formulation for rotations. We explicitly derive the correspondence between this formulation and the dynamics of the individual velocities, thus obtaining a new model of collective motion. In the appropriate overdamped limit we recover the well-known Vicsek model, which dissipates rotational information and does not allow for polarized turns. Although the new model has its most vivid success in describing turning groups, its dynamics is  intrinsically different from previous ones 
in a wide dynamical regime, while reducing to the hydrodynamic description of Toner and Tu  at very large length-scales.
The derived framework is therefore general and it may describe the collective motion of any strongly polarized active matter system.
\end{abstract}

\maketitle 


Flocks of birds represent paradigmatic cases of emergent collective behavior, where short range interactions between individuals lead to nontrivial movement patterns on the large scale\cite{camazine+al_01,couzin+krause_03,giardina_08,sumpter_10,cavagna+giardina_14}.  
Several models have been developed by biologists \cite{aoki_82,reynolds_87,huth_92,couzin+al_02,hemelrijk_10}, physicists \cite{vicsek+al_95,toner+tu_95,gregoire+chate+tu_03,gregoire+chate_04,dorsogna+al_06,huepe+al_07,ginelli+chate_10} and control theorists \cite{krishna_04,tanner+al_07} to describe flocking behavior. Some of these models greatly improved our understanding of collective motion, offering a powerful description of the large scale properties  of active systems \cite{toner+tu_98,ramaswamy_review,vicsek_review,marchetti_review}.
However, recent experiments on real flocks  \cite{attanasi+al_14} revealed new surprising features in the way collective turns are performed that are still unaccounted for, urging for a novel theoretical explanation.


The most conspicuous feature of flocking  is the presence of collective order: flocks are strongly polarized groups, where individuals all move in approximately the same direction. This kind of orientational order can be naturally explained in terms of alignment interactions between the birds velocities and basically all models of flocking include a `social' force term, describing the tendency of individuals to adjust their flight direction to those of neighbors. 
Recently, using inference techniques on real data of flocking birds \cite{bialek+al_12}, it was in fact shown that simple models based on pairwise alignment topological interactions are able to explain the long-range correlations between flight directions and the large degree of coherence exhibited by natural flocks \cite{cavagna+al_10}.

The role of alignment is emphasized in the simplest model of flocking, the Vicsek model \cite{vicsek+al_95}. In the Vicsek model flocking individuals are described as self-propelled particles of constant speed, where the velocity (i.e. the flight direction) of each particle is updated from one time step to the next by computing the average direction of motion of neighbors. Several variants of the model have been analyzed in the literature, where additional cohesive terms are added \cite{toner+tu_98,couzin+al_02,gregoire+chate+tu_03,gregoire+chate_04}, and interactions are chosen topologically rather than metrically (as in the original version) \cite{ballerini+al_08a,ginelli+chate_10}. In all these cases the structure of the dynamical equations always remains the same, where alignment forces due to neighbors directly act on the individual velocity. In other terms, the velocity evolves according to a first order Langevin equation, its time increments being determined by the social force and random noise. This kind of dynamics generates long-range order at low noise and scale-free correlations in the polarized phase, both experimentally observed features of natural flocking. Additionally, at very large scales all these models are well described by the hydrodynamic theory of flocking, whose intriguing predictions of giant density fluctuations, propagating sound modes and anomalous diffusion have been observed in a variety of active inanimate and animate fluids \cite{toner+al_98,toner+tu_98,ramaswamy_review,marchetti_review}.

\begin{figure}
  \centering
\includegraphics[width=.7 \columnwidth]{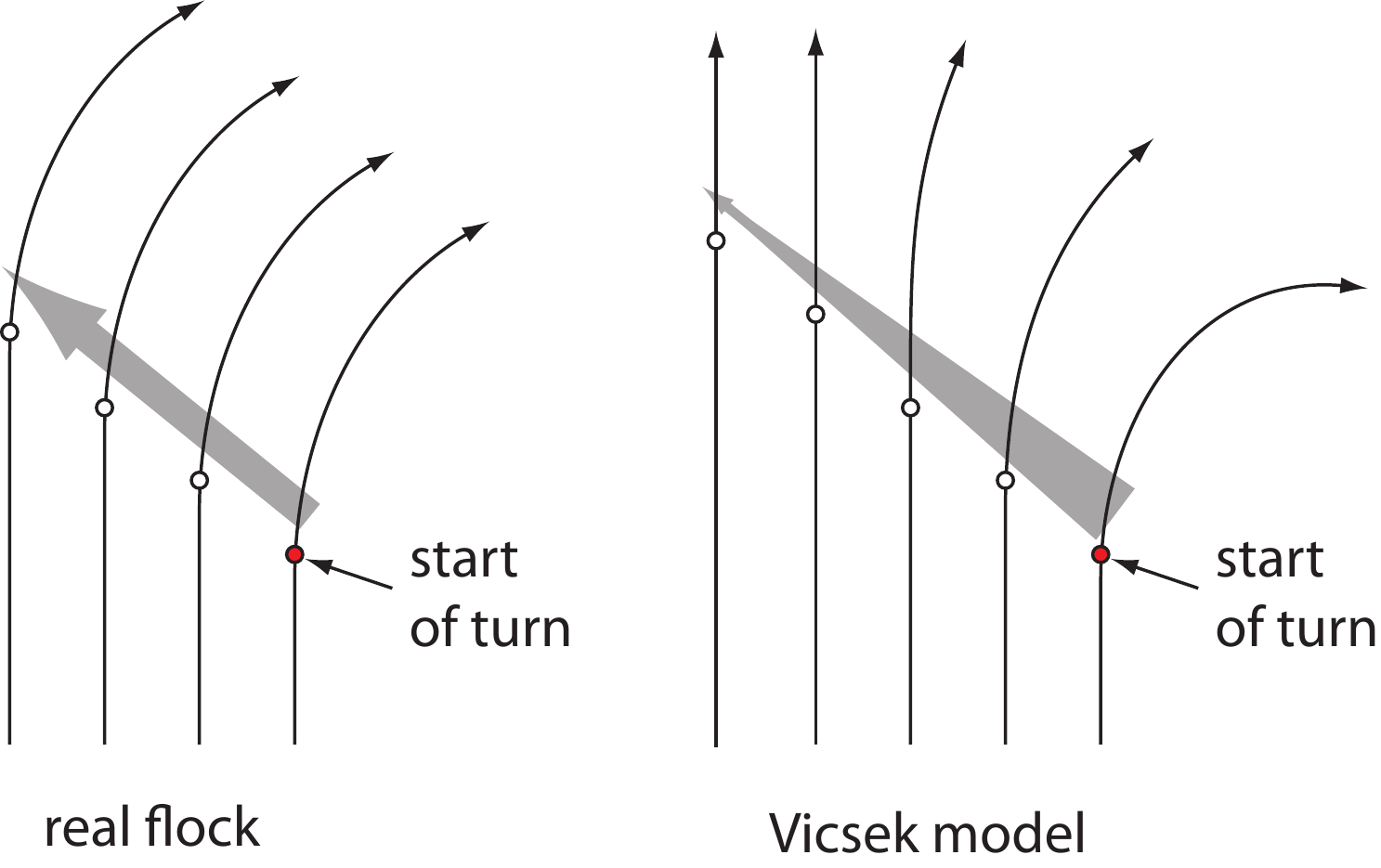}
\caption{Schematic representation of turn propagation in real and Vicsek flocks. {\bf a}. In real flocks when a bird starts a turn the turning information is propagated unattenuated through the whole group and the flock as a whole performs a collective turn. {\bf b.} In the Vicsek model the turning information is damped and does not propagate. The representation concerns scales (group sizes and time) where collective turns occur. In the hydrodynamic regime the Vicsek model exhibits propagating sound modes - see Sec.\ref{sec:transition}.
}
\label{fig:cartoonpropa}
\end{figure}

However, the Vicsek model does not correctly describe the way collective turns occur in natural flocks. Recent experiments on starling flocks \cite{attanasi+al_14} show that collective turns involve linear propagation of purely directional information (the direction of motion of the birds) and curvature (of the individual trajectories), without any local change in density. Turns start locally in space (one single individual starts the turn) and occur very quickly, the turning front propagating with a speed that does not depend on the density of the flock. These  `turning modes' travel extremely fast and undamped through the group, with a linear dispersion law  analogous to superfluid transport \cite{attanasi+al_14}. This is different from what happens in the Vicsek model: the hydrodynamic theory of flocking, that describes the long-wavelength behavior of the Vicsek  model, predicts propagating  sound modes that couple orientational and density disturbances 
\cite{toner+al_98,toner+tu_98}. These are not, however, the kind of modes observed in experiments, where density does not  play any role and curvature propagates together with direction. In fact, turns typically occur in finite groups and on short scales, outside the hydrodynamic regime described by \cite{toner+al_98,toner+tu_98}. On such scales the Vicsek model does not account for directional propagation:  if a directional disturbance is created in the system, such as a few individuals willing to turn and change direction of motion, the turning information is transmitted diffusively and attenuated, i.e. `turning' modes are non-propagating overdamped modes (see Fig.~\ref{fig:cartoonpropa} and sec. V).

In Ref. \cite{attanasi+al_14} some of us showed that there are two crucial and connected ingredients missing in the microscopic dynamics of the Vicsek model, that must be taken into account to theoretically reproduce the dispersion law observed in turning flocks: 1) the presence of a conservation law associated to the rotational symmetry of the system; 2) the existence of a behavioral inertia mediating the effect of the social force.  
In this paper we introduce a new microscopic model for flocking where these two ingredients are appropriately incorporated. This model provides a unified general set of equations to describe systems of interacting self-propelled individuals, where all the salient traits of real flocking - not only order and  correlations, but also turns and information propagation - are correctly reproduced. Interestingly, the Vicsek model can be retrieved as the overdamped limit of this more general model. Additionally, in the long wavelength limit the model is well described by the hydrodynamic theory of flocking of Toner and Tu \cite{toner+tu_98} and is therefore consistent with its predictions on the large scale behavior.

Because the derivation of our new model will be somewhat lengthy, we start by stating the final equations. As we shall see, the crucial ingredient of the new dynamics is the conservation of the internal angular momentum, namely of the spin and the introduction of a spin inertia. For this reason we will refer to these new equations as the {\it inertial spin model}:
\bea
\frac{d\vec{v}_i}{dt}&=&\ \frac{1}{\chi} \vec{s}_i \times \vec{v}_i \nonumber\\ 
\frac{d\vec{s}_i}{dt} &=& \vec{v}_i(t) \times  \left [ \frac{J}{v_0^2} \sum_{j} n_{ij} \vec{v}_j -\frac{\eta}{v_0^2}\frac{d\vec{v}_i}{dt} + \frac{\vec{\xi}_i}{v_0}\right ]\label{eq:model}\\
\frac{d\vec{r}_i}{dt}&=&\ \vec{v}_i(t)\nonumber
\eea
with noise correlator
\be
\langle \vec{\xi}_i(t) \cdot\vec{\xi}_j(t')\rangle = (2d) \;\eta\; T\; \delta_{ij} \, \delta(t-t') \ .
\ee
In these equations $\vec{r}_i$ represents the position of particle $i$, $\vec{v}_i$ its velocity  (with fixed speed $\vert\vec {v}_i\vert=v_0$), and $\vec{s}_i$ is a new variable associated to each particle, representing a generalized momentum (the spin), connected to the instantaneous curvature of the particle's trajectory. The parameter $\chi$ is a generalized moment of inertia, $\eta$ a friction coefficient, and $J$ the strength of the alignment force to neighbors. The connectivity matrix $n_{ij}$ describes who is the neighbour of whom. Finally $T$ is a generalized temperature.

These equations might look unfamiliar and difficult to interpret, but we will show that they are quite natural. Most of the paper is devoted to explaining how the equations  can be derived from general symmetry considerations, what is their meaning, what are the interesting limiting cases (the Vicsek model and the deterministic limit), and what are the predictions of the model for information propagation in the ordered phase. We hope the reader will go through this whole exercise, but for those less interested in the details, the last section of the paper presents in a self-contained way numerical simulations of model (\ref{eq:model}) and illustrates its behavior.

\section{Setting up the stage}

Our aim in this work is to introduce a model of flocking able to describe collective turning and information propagation in flocks. Since these phenomena occur on finite scales of space and time, we need to go beyond hydrodynamics. As a consequence, the universality of the long length scale behavior is lost, and a degree of arbitrariness arises in the way we chose the terms of the microscopic dynamical equations. One way to address this issue is to propose a model and check a posteriori its validity. The reader can look at Eqs. ~(\ref{eq:model}) as our guess, and then directly skip to the last section, where we show with numerical simulations that model (\ref{eq:model}) correctly describes the experimental results of \cite{attanasi+al_14}.

In the following sections we will  follow a longer route, and show that it is possible to construct a microscopic model of collective motion starting from effective variables, symmetries, conservation laws and Hamiltonian dynamics. 
This may seem a bit formal when we think about systems as complex and highly dissipative as real flocks.
However, our point will be that all the nontrivial dissipative nature of animals' active motion can be captured by an effective constraint on their velocity. Once we find the right set of canonical variables enforcing this constraint, we are able to give a Hamiltonian description of flocks. In a second part of the paper, we then generalize this description to include non-Hamiltonian contributions and noise.


In Sec. \ref{sec:circular} we want to make this formal description easier to understand by going back to a very simple (in fact, quite trivial) example, showing that the use of effective variables and symmetries in presence of a constraint is quite standard in physics.

For the sake of simplicity we will assume in this section that bird velocities lie on a plane. Although this is true in most experimental cases \cite{attanasi+al_14}, it is not an essential mathematical condition, and in the next sections we will analyze the fully three-dimensional case. We will also disregard the reciprocal movement of birds (which is very low during real turns \cite{attanasi+al_14}) when  deriving analytically an expression for the dispersion law. Later in Sec. \ref{sec:simulations} we then perform numerical simulations of  model (\ref{eq:model}) where the complete dynamics of positions, velocities and spins is implemented, and network rearrangements are fully taken into account.

\subsection{Circular motion in the external space of position}
\label{sec:circular}
Consider a point on a plane, with coordinate $\vec r$ and linear momentum $\vec p$, subject to the constraint,
\begin{equation}
|\vec r \, |=r_0  \ .
\label{suka}
\end{equation}
The constraint indicates that the point is confined on a circle of radius $r_0$. How is the circular constraint implemented in practice? We do not know, but probably in a complicated way. We know there must be forces acting on the particle that are hard to describe explicitly.  We also know that the standard canonical variables $(\vec r,\vec p)$ are not very convenient to describe the system. The standard thing to do is to introduce a new coordinate that automatically enforces constraint (\ref{suka}). This coordinate is the polar angle $\theta$, whose relation to the coordinate of the particle is simply
\begin{equation}
\vec{r}=r_0 \exp{(i\theta)}  \ .
\end{equation}
The variable $\theta$ parametrizes the rotation in the (external) space of coordinates. The momentum conjugated to $\theta$ is the generator of this rotation \cite{goldstein_80,fetter+walecka_12}, namely the orbital angular momentum, $l$. Hence, $(\theta, l)$ are a pair of canonical conjugated variables and, in absence of forces other than the ones enforcing the constraint, the Hamiltonian is given by
\begin{equation}
H =  \frac{l^2}{2I} \ ,
\end{equation}
where $I$ is the moment of inertia -- the resistance of the point to change its angular momentum. The Hamilton equations of motion are
\begin{eqnarray}
\frac{d \theta}{dt} &=& \{\theta,H\}= \frac{\partial H}{\partial l} = l/I  \ ,
\label{lapo}
\\
\frac{d l}{dt} &=& \{l,H\}   = - \frac{\partial H}{\partial \theta} = 0  \ ,
\label{frank}
\end{eqnarray}
where the Poisson bracket is defined as
\begin{equation}
\{A,B\} = \frac{\partial A}{\partial \theta} \frac{\partial B}{\partial l} - \frac{\partial A}{\partial l}\frac{\partial B}{\partial \theta}  \ .
\end{equation}
The fact that the angular momentum $l$ is the generator of the rotation parametrized by $\theta$ is expressed by the relation,
\begin{equation}
\frac{d A}{d\theta} = \{A, l\} \ .
\end{equation}
The rotational symmetry of the problem ensures (through Noether's theorem) that the generator of the rotation, $l$, is conserved, so that the particle performs rotational uniform motion with constant angular velocity, $\dot \theta$. However, if a dissipative term proportional to $\dot \theta$ were present in (\ref{frank}) the particle would eventually stop rotating, unless there is some external injection of angular momentum.

The message of this repetition of classical mechanics is that, regardless of the complexity by which the constraint is enforced in practice (\ref{suka}), once we introduce the right canonical variables we can completely forget about these complications. Additionally, the symmetry of the Hamiltonian with respect to the transformation parametrized by the effective coordinate ensures the conservation of the generator of the transformation itself, that is the momentum.

\subsection{Circular motion in the internal space of velocity}

Flocks are out-of-equilibrium systems, where energy is continuously injected and dissipated at the individual level.
For this reason, linear momentum is not conserved and it is not possible to use a Hamiltonian description in the standard canonical variables (positions and velocities). However, the only crucial consequence of this very complicated aerodynamics and energetics of bird motion (including the interaction with the surrounding medium) is that individuals move at approximately constant speed. Experimental findings show that fluctuations in the speed are indeed very small, both during straight flight \cite{cavagna+al_10} and during turns \cite{attanasi+al_14}. The simplest way to describe the active nature of the system is therefore to model individuals as self-propelled particles moving at constant speed, as in the original Vicsek model,
\begin{equation}
|\vec v \, |=v_0  \ .
\label{suka2}
\end{equation}
This constraint on the velocity of a single bird is analogous to the constraint on the position of standard circular motion which we have seen above, equation (\ref{suka}).
Hence, as in that case, we must find a generalized coordinate which automatically enforces this constraint. This coordinate is the phase, $\varphi$, namely the orientation of the velocity on the plane,
\begin{equation}
\vec{v}=v_0 \exp{(i\varphi)}  \ .
\end{equation}
Note that the phase $\varphi$ is the parameter of the rotation in the {\it internal} space of velocity, while $\theta$ was the parameter of the rotation in the {\it external} space of positions. This difference is crucial: the generator of the rotation parametrized by $\varphi$ is {\it not} the orbital angular momentum $l$, but rather the internal angular momentum, or spin, $s$. The variables $(\varphi,s)$ are the canonically conjugated pair enforcing constraint (\ref{suka2}), exactly as $(\theta, l)$ are those enforcing constraint (\ref{suka}). For this reason, the single-particle Hamiltonian in absence of forces other than those enforcing the constraint is given by
\begin{equation}
H =  \frac{s^2}{2\chi} \ ,
\label{lubecca}
\end{equation}
where $\chi$ is a generalized moment of inertia quantifying the resistance of a bird to a change of its spin. The equations of motions are
\begin{eqnarray}
\frac{d \varphi}{dt} &=& \{\varphi,H\}= \frac{\partial H}{\partial s} = s/\chi  \ ,
\nonumber
\\
\frac{d s}{dt} &=& \{s,H\}   = - \frac{\partial H}{\partial \varphi} = 0  \ .
\label{antoni}
\end{eqnarray}
The symmetry of the Hamiltonian under rotations in the internal space ensures that the spin is conserved. Hence, this motion is simply one where the velocity rotates at constant angular velocity, $\dot \varphi = s/\chi$. Here we are considering the idealized case where noise and dissipation are absent in the system, so that the spin is fully conserved. Later on we will introduce dissipation, which causes eventually the spin, and thus the angular velocity, to decay to zero in absence of external injection of spin.

\subsection{From one to many birds}

To describe flocks we must generalize Hamiltonian (\ref{antoni}) to the case of many interacting particles, each one with its phase and spin, $(\varphi_i, s_i)$.
The obvious generalization is the following \cite{attanasi+al_14}:
\be
H=V(\{\varphi_i\})+\sum_i \frac{s_i^2}{2\chi}\label{eq:hamiltonian} \ .
\ee
In this expression, $V(\{\varphi_i\})$ is a potential describing the interaction between particles.
The Hamilton equations of motions are
\begin{eqnarray}
\frac{d \varphi_i}{d t} &=&\frac{\partial H}{\partial s_i}=\frac{s_i}{\chi}
\nonumber
  \\
\frac{d s_i}{d t} &=&-\frac{\partial H}{\partial \varphi_i}=-\frac{\partial V}{\partial \varphi_i}  \ .
\label{eq:hamilton}
\end{eqnarray}
The only thing we need to know about the interaction potential $V$ is that it is rotationally symmetric: because all directions of motion are {\it a priori} equally likely at the individual and collective level, it is reasonable to expect that interactions between individuals will respect this symmetry. This means that $V$ (and therefore the Hamiltonian) is invariant under a global rotation of all the velocity vectors $\vec{v}_i$, that is under the transformation $\varphi_i \to \varphi_i +\delta\varphi$. Thanks to Noether's theorem, this global symmetry implies a global conservation law. Although the individual spin $s_i$ is now {\it not} constant, due to the force $\partial V/\partial \varphi_i$, the {\it total} spin, $S=\sum_i s_i$, is conserved. This is clear from Hamilton's second equation,
\begin{equation}
\frac{dS}{d t} =- \sum_i \frac{\partial V}{\partial \varphi_i} = 0  \  ,
\label{nume}
\end{equation}
where the r.h.s. vanishes as a consequence of the symmetry, $V(\{\varphi_i\})=V(\{\varphi_i + \delta\varphi\})$.
We will see later on that in the continuum space limit, equation (\ref{nume}) becomes a continuity equation for the spin field. We will also see that this conservation law is the crucial mathematical ingredient that allows for the propagation of undamped turns in natural flocks \cite{attanasi+al_14}.

It is very important to understand that in the case of many particles, a uniform rotation of the polar angle $\theta$ and a uniform rotation of the phase $\varphi$ give two completely different results. The $\theta$-rotation, generated by the orbital angular momentum $l$, acts on the positions of the points and therefore it gives rise to {\it parallel paths} trajectories: these all have the same origin as a centre of rotation, but different radii of curvature. On the other hand, the $\varphi$-rotation, generated by the spin $s$, acts on the velocities of the points, giving rise to {\it equal radius} trajectories: these have different centres of rotation, but the same radius of curvature. Turning flocks of birds are known to move along equal radius trajectories \cite{heppner_92, attanasi+al_14},
as this is the only way to keep cohesion at constant speed. This experimental observation confirms that phase and spin are indeed the correct canonical variables.

Let us summarize this introductory Section. We describe birds' motion through a set of effective conjugated variables, the phase $\varphi_i$ and the spin $s_i$, that automatically enforce the constraint of constant speed, $|\vec v_i \, |=v_0$. This allows us to disregard the highly complex and dissipative mechanism enforcing this constraint and to switch to a Hamiltonian description. Once the speed is constant, the only transformation that the velocity can undergo is a rotation, which is parametrized by the phase. The generator of this rotation is the spin (internal angular momentum), so that in presence of rotationally symmetric forces the total spin of the system is conserved.
The way real birds turn is of course complicated, involving the details of flight aerodynamics. However, as long as birds are able to control their motion keeping their speed fixed and performing equal radius turning - as they do - they can be described in a simple way within this framework.

\section{Deterministic equations for the velocity and the spin}
\label{sec:suka}

We have written the dynamical equations for the phase and the spin, which are the canonical variables, but not for the velocity. However, in order to actually run a simulation of a system of self-propelled particles, we need to know how the velocities evolve in time. Because the velocity is {\it not} a canonical variable, the update equations for $\vec v_i$ will be nontrivial. Moreover, we need to study the fully three-dimensional case, in order to have a model as general as possible. In this Section we will write the dynamical equations for the velocity, in absence of noise and dissipation, which amounts to describing a deterministic flock. Noise and dissipation will be introduced later on.

\subsection{From the phase to the velocity}

In the context of Hamiltonian dynamics, there is a standard method to retrieve the dynamical evolution of any observable once the Hamiltonian is given in terms of canonical variables (here phase and spin). This method is to calculate  the Poisson brackets with the Hamiltonian, which is the time-evolution operator.
In the simple planar case discussed in the previous section, we have ${v}_i=\exp({i \varphi_i})$, hence
\be
\frac{dv_i}{dt} = \{v_i,H\}=\frac{\partial v_i}{\partial \varphi_i}\frac{\partial H}{\partial s_i}
-\frac{\partial v_i}{\partial s_i}\frac{\partial H}{\partial \varphi_i}  
= \mathrm{i} \, v_i \frac{s_i}{\chi}  \ .
\label{naso}
\ee
In the planar case the velocity is a two dimensional vector on the $(x,y)$ plane, whereas the spin (the generator of the rotations of $v$ in the plane) can be seen as a vector along the $z$ direction orthogonal to the plane. In this vectorial description, the term $\mathrm{i}\, v_i s_i$ at the r.h.s of (\ref{naso}) is a vector orthogonal to both $v_i$ and $s_i$, hence it is  simply the cross (vector) product between $v_i$ and $s_i$. We therefore can write (with a slight abuse of notation)
\be
\frac{d \vec v_i}{dt} = \frac{1}{\chi} \vec s_i \times \vec v_i  \ .
\label{olga}
\ee
The derivation of equation (\ref{olga}) can be generalized to the case of a fully three-dimensional velocity. In this case we must introduce three phases, $\{\varphi^a_i\}$ ($a=x,y,z$), each one parametrizing the rotation around a different cartesian axis. Accordingly, the generator of the symmetry, i.e. the spin, becomes a full three-dimensional vector, $\vec{s}_i=\{s_i^a\}$. With this notation, for example, $\varphi_z$ is the angle parametrizing rotations around the $z$ axis (i.e. acting in the $xy$ plane) and $s_z$ is the corresponding spin (oriented along $z$). Rotations around the $z$ axis  leave the $z$ component of $\vec{v}$ unaltered while effectively rotating the projection of $\vec{v}$ in the plane orthogonal to $z$. Thus $\varphi_z$ plays the role of a polar coordinate in the $xy$ plane, and it is equivalent to $\varphi$ in the planar case discussed in the previous section. We have, 
\begin{eqnarray}
v_y&=&\sqrt{v_0^2-v_z^2} \sin(\varphi_z)
\\ 
v_x&=&\sqrt{v_0^2-v_z^2} \cos(\varphi_z) \ .
\end{eqnarray}
Analogous relations hold connecting $\varphi_x$ and $\varphi_y$ to the components of $\vec v$.
We can now explicitly compute the equations for the velocities by using Poisson's brackets,
\bea
\frac{dv_i^a}{dt} &=& \{v_i^a,H\}=\sum_c \frac{\partial v_i^a}{\partial \varphi_i^c}\frac{\partial H}{\partial s_i^c}
-\frac{\partial v_i^a}{\partial s_i^c}\frac{\partial H}{\partial \varphi_i^c}
= \epsilon_{abc} v_i^b \frac{s_i^c}{\chi}
\nonumber\\
\frac{ds_i^a}{dt} &=& \{s_i^a,H\}=
\sum_c \frac{\partial s_i^a}{\partial \varphi_i^c}\frac{\partial H}{\partial s_i^c}
- \frac{\partial s_i^a}{\partial s_i^c}\frac{\partial H}{\partial \varphi_i^c}
=  \epsilon_{abc} v_i^c \frac{\partial H}{\partial v^b_i}\nonumber\\
\label{eq:poisson}
\eea
where $\epsilon_{abc}$ is the fully antisymmetric Levi-Civita symbol, and we exploited the fact that $\partial v_i^a/\partial \varphi_i^b=-\epsilon_{abc} v_i^c$. We can rewrite Eqs.~(\ref{eq:poisson}) in a more compact form using vectorial notation, and finally get the set of dynamical equations of the inertial spin model (deterministic case),
\bea
\frac{d\vec{v}_i}{dt}&=&  -\vec{v}_i \times \frac{\delta H}{\delta \vec{s}_i}=\frac{1}{\chi} \vec{s}_i  \times \vec{v}_i \label{eq:det1}\\
\frac{d\vec{s}_i}{dt} &=& - \vec{v}_i \times \frac{\delta H}{\delta \vec{v}_i}=\vec{v}_i\times \vec{F}_i\label{eq:det2}\\
\frac{d\vec{r}_i}{dt}&=& \vec{v}_i  \ ,
\label{eq:det3}
\eea
where we have also added the kinematic relationship between the velocity $\vec{v_i}$ of a particle and its position in space $\vec{r}_i$.  We note that these equations are formally analogous to what one would get for rotations and circular motion in real space, with the external variables  $\vec{r}$ and $\vec{l}$  playing the role of $\vec{v}$ and $\vec{s}$. In this last case, though, equation (\ref{eq:det3}) would of course not be present.

In (\ref{eq:det1}-\ref{eq:det3}) $H$ is the same Hamiltonian as in (\ref{eq:hamiltonian}) (but written in terms of velocities rather than phases) and  $\vec{F}_i=-\delta H/\delta\vec{v}_i$ is  therefore the force acting on particle $i$ due to interactions with other particles. 
For alignment interactions we have
\be
H(\vec{v},\vec{s})= -\frac{J}{2v_0^2}\sum_{ij} n_{ij} \vec{v}_i\cdot\vec{v}_j+ \frac{1}{2\chi} \sum_i s_i^2 \ ,
\label{eq:hamiltonian-v}
\ee
where $n_{ij}$ is the connectivity matrix (being $1$ if $j$ is a particle interacting with $i$ and $0$ otherwise), and $J$ is the alignment strength.  Therefore $\vec{F}_i=(J/v_0^2) \sum_j n_{ij} \vec{v}_j$. Note that, contrary to Eqs.~(\ref{eq:hamilton}), Eqs.(\ref{eq:det1}-\ref{eq:det3}) do not have a Hamiltonian form, precisely because $(\vec{v}_i,\vec{s_i})$ are not canonical variables ($\vec{s}_i$ is the conjugated moment to the phase $\vec \varphi_i$, not to the velocity vector $\vec{v}_i$). Still, 
they retain a pseudo-Hamiltonian structure where the derivatives of the Hamiltonian are combined with a vectorial product. Thanks to this structure, the speed of the individual particles is automatically conserved by the dynamics. Besides, the equations also conserve $\vec{s}_i\cdot\vec{v}_i$, which we fix equal to zero (the only solution in absence of forces). Finally, the Hamiltonian $H$ itself is conserved by these equations.

\begin{figure}
 \centering
 \includegraphics[width=0.5\columnwidth]{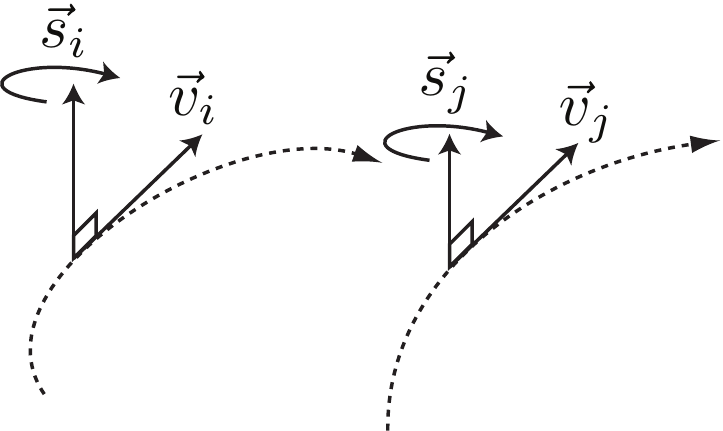}
\caption{Schematic representation of velocity, spin, and trajectory for two particles.
}
\label{fig:cartoonspin}
\end{figure}

We can now explore how Eqs.(\ref{eq:det1}-\ref{eq:det3}) determine the motion of particles in space (see Fig.~\ref{fig:cartoonspin}). When the spin $\vec{s}_i$ is equal to zero, the direction of motion of particle $i$ always remains the same and the particle therefore follows a straight  path in real space. When $\vec{s}_i$ is different from zero but constant,  the flight direction obeys uniform circular motion and the particle performs a turn in real space with a constant radius of curvature $R\sim v_0 \chi/|\vec{s}_i|$.  Hence,  the spin $\vec{s}_i$ has a clear kinematic meaning being related to the instantaneous curvature $\kappa=1/R$ of the trajectory.
When there are forces acting on the particle the local spin/curvature $\vec{s}_i$ changes in time, inducing a variation in the rate of direction changes. The paths followed by the particles in real space depend on the instantaneous realization of the forces. However, the important point is that - whatever these forces are - in our new model they act on the spin $\vec{s}_i$ and not directly on the velocity, $\vec v_i$. In other terms, forces cannot change the direction of motion abruptly, but there is an inertial effect mediated by $\vec{s}_i$.  In this respect, we see that the generalized inertia $\chi$ measures the resistance of the particle to change the instantaneous radius of curvature of its trajectory. 
In the case of birds, $\chi$ can be explicitly related to the resistance of the bird to change its banking angle (see \cite{attanasi+al_14}).

\subsection{Spin conservation and information propagation}
Another crucial consequence of the presence of a global continuous symmetry is that the generator of the symmetry is a constant of motion of the dynamics. In our case, this implies that the global spin $\vec{S}=\sum_i\vec{s}_i$ is conserved. Interestingly, for a polarized flock this conservation law also has  local implications (i.e. at the level of individual particles): if a strong misalignment occurs between a particle $i$ and its neighbors, this causes an excess of curvature and  an excitation of the local spin $\vec{s}_i$; since the global spin is a conserved quantity, the local excitation cannot be re-absorbed or dissipated, but must be carried away. For this reason we expect propagating modes in this system. 

To better understand how this occurs, let us consider a polarized flock of particles all moving approximately in the same direction. We stick  to the simpler planar case for convenience, but the argument is general. For an ordered flock, the rotational symmetry is spontaneously broken since a specific global direction of motion is selected by the system. Still, a residual symmetry remains in the space orthogonal to this direction: we can choose the phase $\varphi_i$ of each particle as the angle with respect to the global flock velocity, and use Eqs.(\ref{eq:hamilton}) to describe the evolution of such phases. 

Since the flock is ordered, phases are small and the potential can be written as $V=J/4\sum_{ij} n_{ij} [\varphi_i-\varphi_j]^2$. If phase variations are smooth and we look at distances larger than the typical distance $a$ between interacting neighbors, we can consider the continuum limit in space $\varphi_i(t)\to \varphi(\vec{r},t)$, $s_i(t)\to s(\vec{r},t)$. The phase variation between a particle and its neighbors $\sum_j n_{ij} [\varphi_i-\varphi_j]^2$ can be conveniently approximated by $ J n_c a^2 [\nabla \varphi]^2$, where $n_c=(1/N)\sum_i\sum_j n_{ij}$ is the average number of interacting neighbors. Then, Eqs.(\ref{eq:hamilton}) become
\begin{eqnarray} 
\frac{\partial\varphi}{\partial t}&=&\frac{s}{\chi} \\
\frac{\partial s}{\partial t}&=&n_ca^2 J \nabla^2 \varphi  \ .
\label{eq:con1}
\end{eqnarray}
Interestingly, the second of these equations can be written as a continuity equation for the field $s(\vec{r},t)$,
\be
\frac{\partial s}{\partial t}-\nabla\cdot \vec{j}=0 \ ,
\label{eq:con2}
\ee
with current $\vec{j}=n_ca^2 J \vec \nabla \varphi$. This conservation law describing spin flow is precisely the local counterpart of global spin conservation that we mentioned above. If we now take a second derivative of $\varphi$ with respect to time, we immediately get d'Alembert's equation, 
\be
\chi\frac{\partial ^2 \varphi}{\partial t^2}=J n_c a^2 \nabla^2\varphi \ .
\label{eq:con3}
\ee
This equation can be easily solved in Fourier space to get the dispersion relation, namely the law describing how directional information travels through the system. We find
\be
\omega=c_s k  \ ,
\label{eq:linear}
\ee
with $c_s=\sqrt{Jn_ca^2/\chi}$. A real value for the frequency $\omega$ corresponds to propagating modes: what we find is therefore that a deterministic flock exhibits undamped propagating modes of the phase. Besides, the dispersion law is linear, meaning that propagating modes travel at a well defined speed $c_s$, which is a function of the alignment strength and the generalized inertia. Linear propagation is reminiscent  of sound propagation in a medium. Here, however, the modes that we are looking at are not related to density fluctuations, but to phase fluctuations. In fact the same equations for the phase we are looking at, and the same dispersion law, would also hold for a fixed network of particles, e.g. a regular lattice. Phase propagating modes mean that if in a flock  a particle starts turning, this change will affect through the alignment term the spin/curvature of nearby particles, which will start turning themselves. The information will travel with a speed $c_s$ which is larger the more ordered the flock, and the whole system will perform a collective turn. As we shall discuss, this is similar to what happens in natural flocks of birds, indicating that far from being an idealized abstract description of collective motion, the deterministic model introduced so far has some crucial ingredients occurring in real systems.

The field equations we have written for $\varphi(\vec{r},t)$ and $s(\vec{r},t)$ in this Section make one important approximation, namely that the interaction matrix $n_{ij}$ is fixed in time. Without this approximation we cannot use the continuous Laplacian, nor obtain d'Alembert's equation. This condition is, in general, violated, and this is what marks the difference between a lattice model and an off-lattice model of moving particles, where $n_{ij}= n_{ij}(t)$ depends on time. However, if the time scale of propagation in d'Alembert's equation (\ref{eq:con3}) is shorter than the typical time scale of reshuffling of the network $n_{ij}$, then the equations are justified. This is indeed the case in natural flocks of starlings \cite{attanasi+al_14}, where phenomena like collective turns happen on a time scale indeed significantly shorter than the time needed for an individual to change its interacting neighbours. Nevertheless, at the end of this work we will run numerical simulations of the full model with varying $n_{ij}(t)$ and show that when $c_s$ is large enough, the propagation phenomena are quantitatively in agreement with the field equations we have written above.

\subsection{Dynamical universality class}
The relationship between symmetries of the system, equations of motions and conservation laws  is a well known feature of Hamiltonian dynamics and a particular manifestation of Noether's theorem. The presence of a continuous symmetry implies the existence of a conserved quantity, and the concomitant linear propagation of local disturbances. What is novel in our analysis is that we have studied these properties in the context of the dynamics of flight directions of flocking particles.

Interestingly, the same structure of equations and dispersion law that we have introduced for flocks also hold in other kinds of systems, where a non-conserved order parameter (the flight direction in the case of flocks) co-evolve with a conserved quantity (e.g. the spin/curvature). In fact, Eqs. (\ref{eq:con1})(\ref{eq:con2}) are formally identical and belong to the same dynamical universality class  as the planar ferromagnet \cite{matsubara+matsuda_56b} and superfluid Helium \cite{helium_47,halperin_69} (model F in the Halperin-Hohenberg classification \cite{halperin+hohenberg_77}). In liquid Helium, for example, the quantum phase plays the role of the flight direction, the (conserved) superfluid component plays the role of the spin, $\vec{j}$  is the superfluid current, and the propagating wave is the so-called `second sound' \cite{helium_47,halperin_69,sonin_10}. In the same way, the full three-dimensional model Eqs~(\ref{eq:det1}-\ref{eq:det3}) is equivalent to the isotropic antiferromagnet  (model G dynamical universality class), with exactly the same equations, the staggered magnetization playing the role of the flight direction, and the (conserved) total magnetization playing the role of the spin. 

In both these cases (model F and model G of  \cite{halperin+hohenberg_77}), propagating modes and a linear dispersion law are the consequence of the rotational symmetry and of the conservation of the spin deriving from it. This is not a generic feature of spin models. For example, in the case of the isotropic ferromagnet, the conserved field is the order parameter itself and this conservation does {\it not} derive from any symmetry of the Hamiltonian. In this case the linear propagation of the phase is lost, and a dissipative, diffusive dispersion law holds \cite{halperin+hohenberg_77}. Therefore, the essential ingredient of linear spin propagation is not simply the presence of a conserved quantity, but rather the fact that this conserved 
quantity must be the generator of a spontaneously broken continuous symmetry. This condition is the true hallmark of superfluid transport, namely of the undamped and linear propagation of the spin. In the case of Helium, the spin is mapped onto the superfluid component, which propagates through second sound \cite{halperin+hohenberg_77}, whereas in the isotropic antiferromagnet one has actual spin superfluid propagation \cite{halperin+hohenberg_77}. In flocks, spin carries the turning information.

\section{A new model of collective motion}

The deterministic inertial spin model we have introduced in the previous Section conserves spin (through the continuity equation), but it does so a bit too effectively, as in absence of forces the angular velocity remains forever constant. Here we want to cure this anomaly.
An example may help: although linear momentum conservation and inertia is essential to propagate standard sound in a crystal, we certainly do not expect sound to propagate forever and for infinite distance, as some dissipation will eventually dampen the signal. The same must be true for second sound, which is carried by spin conservation. Therefore, we must introduce a dissipative term coupled to $\vec s$. Moreover, the dissipative term will be associated to a noise term, making the model stochastic.

It is important to understand that the dissipation we need to introduce is {\it not} the standard linear one, proportional to $\dot{\vec x}$. Standard linear dissipation would drive the modulus of the velocity to zero, while this cannot happen here: the constraint $|\vec v_i| = v_0$ must be automatically enforced by our equations.  Instead, spin dissipation must account for the natural and rather intuitive fact that in absence of spin injection and of interaction with neighbours, a bird is unlikely to turn forever with constant angular velocity. As we shall see, regulating the strength of this dissipative term will make the difference between propagating and nonpropagating information.

\subsection{The Inertial Spin Model}
As we would do for a physical system, we start from the deterministic structure of the equations (\ref{eq:det1}-\ref{eq:det3}) and add noise and dissipation to the `conservative' forces $\vec{F}_i=-\delta H/\delta\vec{v}_i$. In this way we obtain the final equations of the inertial spin model of collective motion,
\bea
\frac{d\vec{v}_i}{dt}&=&\ \frac{1}{\chi} \vec{s}_i \times \vec{v}_i \label{eq:v}\\ 
\frac{d\vec{s}_i}{dt} &=& \vec{v}_i(t) \times  \left [ \frac{J}{v_0^2} \sum_{j} n_{ij} \vec{v}_j -\frac{\eta}{v_0^2}\frac{d\vec{v}_i}{dt} + \frac{\vec{\xi}_i}{v_0}\right ]\label{eq:s}\\
\frac{d\vec{r}_i}{dt}&=&\ \vec{v}_i(t) \ ,
\label{eq:r}
\eea
with $\vec{v}_i\cdot\vec{s}_i=0$. Here $\eta$ is a generalized viscous coefficient and $\vec{\xi}_i$ is an i.i.d  vectorial noise with variance
\begin{equation}
\langle \vec{\xi}_i(t)\cdot \vec{\xi}_j(t')\rangle = (2 d)\,  \eta\,   T \, \delta_{ij} \delta(t-t') \ , 
\end{equation}
where we have introduced the generalized temperature $T$, in analogy to physical systems. Note that, consistently with our entire description, the constraint $|\vec v_i| = v_0$ is satisfied by these equations.

The inertial spin model (\ref{eq:v}-\ref{eq:r}) represents a full description of the system's dynamics in terms of the variables $(\vec{v}_i, \vec{s}_i)$. Being a first order system it is relatively easy to implement numerically, and we shall discuss results of simulations later in the paper. For the time being, we note that there are four important parameters entering Eqs.(\ref{eq:v}-\ref{eq:r}): the alignment strength $J$, the moment of inertia $\chi$ of the spin, the viscous coefficient $\eta$, and the temperature $T$. The spin inertia, $\chi$, is the new ingredient in the dynamics.

To better understand the effect of dissipation, let us consider the case where there are no social forces acting on the birds and the noise is zero  (i.e. $F_i=0$, $T=0$). We remind that under these conditions, if $\eta=0$ a bird conserves its local spin, i.e. it keeps turning with uniform angular velocity. When $\eta\ne0$, on the other hand, Eqs.~(\ref{eq:v}-\ref{eq:s}) give,
\bea
 \frac{d\vec{v}_i}{dt}&=&\ \frac{1}{\chi} \vec{s}_i \times \vec{v}_i \\
\frac{d\vec{s}_i}{dt}& =& -\frac{\eta}{\chi}\vec{s}_i \ .
\eea
The spin exponentially decays to zero, so that the velocity vector stops rotating. A bird that was originally turning ($\vec{s}_i\ne0$) reduces its spin/curvature until reaching a state of straight motion ($\vec{s}_i=0$). Thus $\eta$ acts as a damping on the spin and plays the role of a rotational dissipation.

\subsection{A closed equation for the velocity}

By taking a further time derivative of Eq.~(\ref{eq:v}) and exploiting Eq.~(\ref{eq:s}), one can get a closed second order equation in the velocities,
\bea
&&\chi \frac{d^2\vec{v}_i}{dt^2}+\chi \frac{\vec{v}_i}{v_0^2}\left(\frac{d\vec{v}_i}{dt}\right)^2 + \eta \frac{d\vec{v}_i}{dt}=
\nonumber\\
&&\phantom{ppppppp}
=\frac{J}{v_0^2}\left (\vec{v}_i\times \sum_j n_{ij} \vec{v}_j\right)\times\vec{v}_i+v_0 \vec{\xi}_i^\perp \ ,
\label{eq:second-order}
\eea
with
\be
\vec{\xi}_i^\perp=\vec{\xi}_i-(\vec{\xi}_i\cdot\frac{\vec{v}_i}{v_0})\frac{\vec{v}_i}{v_0}  \ ,
\ee
and
\be
\langle \vec{\xi}^\perp_i(t)\cdot \vec{\xi}^\perp_j(t')\rangle = 2 (d-1) \, \eta\,   T \,  \delta_{ij} \delta(t-t')  \ .
 \ee
Equation (\ref{eq:second-order}) is a generalized Langevin equation with respect to the velocity vectors. The first term in the l.h.s is an inertial second order term, where  $\chi$ is {\it not} the conventional inertia (i.e. the mass of the particle) but measures the resistance of a particle to change the curvature of its trajectory. The second term at the l.h.s. is a contribution arising from the constraint on the speed, the analogue of what would be the centripetal acceleration for circular motion in $\vec r$ space. Its presence is a signature of the underlying Hamiltonian structure of the equations and is therefore essential to recover the appropriate deterministic limit when $\eta \to 0$. The last term in the l.h.s. is the dissipative term. In the r.h.s. we find the effect of the social force and the noise.

As compared to other second order equations that have been considered in the literature \cite{krishnaprasad_04,szabo_09,gautrais+al_12,hemelrijk_11,sumino+al_12}, in Eq.(\ref{eq:second-order}) the connection between inertial terms and symmetries is automatically implemented. We stress that this is in fact a crucial point, since - as we showed in the previous sections - the way information propagates is strongly influenced by this connection.

\subsection{Overdamped limit: the Vicsek model}

Interestingly, the original Vicsek model happens to be the overdamped limit of the inertial spin model, that is the limit in which the spin inertia becomes negligible compared to the spin dissipation.
Equation (\ref{eq:second-order}) is the full Langevin equation for the velocity, including both inertial terms (proportional to $\chi$) and a dissipative term (proportional to $\eta$). As usual in the Langevin equation \cite{zwanzig,gardiner}, the overdamped limit is obtained by taking the 
limit $\chi/\eta^2 \to 0$, which in our case gives
\be
\eta \frac{d\vec{v}_i}{dt}=\frac{J}{v_0^2}\left (\vec{v}_i\times \sum_j n_{ij} \vec{v}_j\right)\times\vec{v}_i+v_0  \vec{\xi}^\perp_i \ .
\label{vicsek}
\ee
Note that a rescaling of time, $t\to t'=t/\eta$, would allow us to get rid of $\eta$ both in the equation for $\vec v$ and in the noise correlator (this is the reason why the overdamped limit is $\chi/\eta^2 \to 0$, rather than $\chi/\eta \to 0$). This time rescaling is usually taken for granted in the overdamped limit and for this reason $\eta$ is normally set equal to $1$ in this limit.

Equation (\ref{vicsek}) is identical to the Vicsek model (in the continuous time limit version). The double cross product in the r.h.s. has a clear interpretation. Calling $\vec{F_i}=(J/v_0^2)\sum_j n_{ij} \vec{v}_j$ the social force acting on particle $i$, we can exploit the properties of the cross product and write
\bea
(\vec{v_i}\times  \vec{F}_i)\times\vec{v}_i&=& v_0^2[\vec{F}_i-(\vec{F}_i\cdot\vec{v}_i)\vec{v}_i/v_0^2]\nonumber\\
&=&v_0^2 F_i^\perp=J(\sum_j n_{ij} \vec{v}_j)^\perp \ .
\eea 
Only the perpendicular component of the social force contributes to changing the velocity vector, as it should be since the speed is constant due to the constraint. Thus, the double cross product is a formal and clean way to ensure that the norm of $\vec{v}_i$ is conserved during the dynamics. 

\subsection{Steady state distribution of velocity and spin}
\label{pussy}

We have seen that the dynamical equations of the inertial spin model have their deep roots in the existence of a symmetry, and its associated conservation law.  In fact, the rotational symmetry and its spontaneous breaking also have implications when looking at the steady state distribution of velocity and spin. 
In general, one might wonder whether a steady state distribution exists at all in flocks, since the system is completely out of equilibrium. In fact many of the most intriguing features of active systems precisely come from their off-equilibrium nature \cite{marchetti_review}. Still, if we look at a flock  on timescales smaller than the typical swapping time between an individual an its neighbors, then we can consider the velocities and spins as dynamical variables evolving on a fixed interaction network. In this case we can show that the joint distribution probability for velocities and generalized momenta obeys a Fokker-Planck equation \cite{zwanzig,gardiner}. The steady state solution of  such an equation reads
\bea
P(\vec{v},\vec{s}) \sim \ &&\exp{ \left [- \frac{1}{T} H(\vec{v},\vec{s})\right ]}=\label{eq:boltzmann}\\
&&
\exp{ \left[-\frac{1}{T}
\left (
-\frac{J}{2v_0^2}\sum_{ij} n_{ij} \vec{v}_i\cdot\vec{v}_j
+ 
\sum_i\frac{s_i^2}{2\chi}
\right)
\right]
} \ .
\nonumber
\eea
As usual in statistical physics, this probability distribution factorizes between coordinates and momenta, that is velocities and spins. This result tells us that velocities are in fact amenable to a statistical description (on scales where network rearrangements can be disregarded): a measure exists, which can be used to compute expectation values.  This distribution is a Boltzmann distribution, where the ratio between noise and dissipation - what we called temperature - in fact acts as a real temperature in the statistical sense. Additionally, inertia does not enter in the velocity distribution. For this reason, all the static properties of our model (including the velocity correlation functions and the steady-state ordering properties), are the same as the Vicsek model, which is described by the very same configurational part.

If we now look at the explicit form of the Hamiltonian in (\ref{eq:boltzmann}), we see that the velocities are described by a Heisenberg model on a random euclidean lattice (the one realized by the particles positions). In the ordered phase, where the rotational symmetry is spontaneously broken, this distribution generates soft Goldstone modes and long range correlations corresponding to fluctuations of the individual velocities perpendicular to the global velocity of the flock \cite{goldstone}.  
 Long-range correlations in velocity fluctuations have been experimentally observed in natural flocks of birds \cite{cavagna+al_10}. Besides, in \cite{bialek+al_12} using inference techniques on empirical data, we showed that the equal time properties of the velocity fluctuations can be fully described by a Boltzmann distribution with local pairwise alignment interactions. The model we have discussed so far is therefore fully consistent with all experimentally known dynamic {\it and} static features of real flocks of birds.

\section{Information propagation in the ordered phase}
\label{sec:propagation}

Equations (\ref{eq:v}-\ref{eq:r}) give rise at low noise to polarized flocks, much as in the Vicsek model. 
In Section \ref{sec:suka}  we have seen that for a deterministic flock the rotational symmetry and the concomitant conservation law have strong implications on the properties of the ordered phase. Now we would like to understand whether this scenario holds in presence of noise and dissipation.

Let us consider the system in the strongly polarized phase. In this case, the individual velocities are very aligned to each other and the flock as a whole has a non-zero collective direction of motion $\vec{V}$. To fix ideas, let us assume that the group velocity is along the $x$ direction,  $\vec{V} = V \vec n_x$, with $\vec n_z = (1,0,0)$. The individual velocities can be conveniently written in terms of a longitudinal component along the direction of motion, $\vec{n}_x$, and a perpendicular component, i.e. a two dimensional vector $\vec \pi$ in the two-dimensional $(y,z)$ plane,
\be
\vec{v}_i=v_i^L \vec{n}_x+\vec{\pi}_i  \ .
\ee
Because of the large polarization all individual velocities will be very close to $\vec V$, so that  ${\pi}_i^2 \ll1$ 
and
\be
v_i^L=\sqrt{v_0^2-{\pi}_i^2}\sim v_0[1-{\pi}_i^2/(2v_0^2)] \ . 
\label{zero}
\ee
Using the phases defined in the previous sections, we can write
\bea
\pi_y &=& v_0 \sin(\varphi_z) \sim v_0\, \varphi_z \ ,
\label{anza}
\\
\pi_z &=& v_0 \sin(\varphi_y) \sim v_0 \, \varphi_y \ .
\label{unza}
\eea
However confusing these relations may seem, they are correct: a nonzero $y$ component of $\vec \pi$ is obtained by rotating $\vec v$ around the $z$ axis, a rotation parametrized by $\varphi_z$. Similarly, a nonzero $\pi_z$ is generated by a rotation parametrized by $\varphi_y$. The crucial point is that, once we have spontaneous symmetry breaking, only {\it two} relevant phases are left.
By using (\ref{zero}-\ref{unza}) into equation (\ref{eq:second-order}) and expanding at the first order (spin wave expansion), we obtain for each one of the two phases, $\varphi_z$ and $\varphi_y$, the same equation,
\be
\chi \frac{d^2{\varphi}_i}{dt^2}=J \sum_j \Lambda_{ij} {\varphi}_j\ - \eta \frac{d{\varphi}_{i}}{dt}+\xi_i^\perp \ ,
\label{eq:second-orderphi}
\ee
where $\Lambda_{ij}=-n_{ij}+\delta_{ij}\sum_k n_{ik}$ is the discrete Laplacian.

In general, solving equation (\ref{eq:second-orderphi}) is not an easy task, because the dynamics of the $\vec{\varphi}$'s is coupled in a non-trivial way with the movement  of the particles in space, i.e. the network $n_{ij}$ changes with time. Hydrodynamic theories of flocking address this problem by coarse-graining the microscopic dynamics and looking at the large-scale long-time behavior. Here, however, we are interested in addressing the dynamics of the system on shorter timescales. We can therefore assume that $n_{ij}$ does not change significantly, as indeed happens in flocks during collective turns \cite{attanasi+al_14}. Later in Sec.~\ref{sec:simulations} we perform numerical simulations of model (\ref{eq:model}) where network rearrangements on the scale of turns are fully taken into account.

For a fixed interaction network, the equation for $\varphi$ can be solved analytically to obtain the dispersion relation.
To do this,  we must choose a diagonal representation both in space and time, i.e. we must diagonalize the Laplacian $\Lambda$, write the equations in terms of its eigenmodes, and Fourier transform with respect to time. This can be done exactly numerically. Yet a simple analytical approximation gives us some intuition about the nature of the dispersion relation. To do this we can proceed as in the previous section: if we look at spatial scales larger than the nearest-neighbor distances, we can approximate the discrete Laplacian with its continuous counterpart, i.e. $J \sum_j \Lambda_{ij}  \to J n_ca^2 \nabla^2$ (where, as before,  $a$ is the typical distance of the interacting neighbors and $n_c$ their number).  In this case we can write
\be
\chi\frac{\partial^2 \varphi}{\partial t^2}=J n_c a^2 \nabla^2\varphi- \eta \frac{\partial \varphi}{\partial t}+ \xi^\perp \ .
\label{eq:continuous}
\ee
The propagator (Green function) of this differential equation can be easily computed in Fourier space giving the following dispersion law
\be
\chi \omega^2-i\eta\omega-Jn_ca^2 k^2 =0  \ .
\ee
This dispersion law tells us how local disturbances in the flight direction of a particle affect the rest of the flock. If a bird/particle changes its flight direction, will this perturbation propagate through the entire group or be damped? The answer to this question depends on the values of the parameters entering the equations, and in particular on the balance between dissipation and inertia. 

In the deterministic limit $\eta\to 0$ we get  the linear dispersion  law $\omega=c_s k$ discussed in the previous sections with
\be
c_s=\sqrt{\frac{Jn_ca^2}{\chi}} \ .
\label{eq:speed1}
\ee
This is the  perfectly undamped case of propagating orientational modes. At the opposite end of the spectrum there is the over-damped limit, $\chi/\eta^2\to 0$, which, as we have seen, corresponds to the Vicsek model. In this case  $\omega$ is purely imaginary with a quadratic diffusive dispersion law, $\omega = i (Jn_ca^2/\eta) k^2$. We conclude that in the Vicsek case there is no propagation, but pure exponential damping. We will see through numerical simulations that this is indeed what happens: in the Vicsek model it is not possible to locally initiate a turn that propagates to the entire flock.

In the general case, where both dissipation and inertia are different from zero, we obtain
\be
\omega = i/\tau \pm \omega_0 \sqrt{1- k_0^2/k^2} \ ,
\label{eq:omega}
\ee
with 
\be
\omega_0\equiv  c_s k \quad , \quad
k_0 \equiv \frac{\eta}{2\sqrt{Jn_ca^2\chi}} \quad ,\quad  \tau \equiv 2\chi/\eta \ .
\label{eq:paradiss}
\ee
Here $\omega_0$ is the zero dissipation frequency, while $k_0$ and $\tau$ are the two relevant scales, respectively in wave number and time, related to the effect of dissipation. With zero dissipation, we get $k_0=0$, $\tau=\infty$ and $\omega = \omega_0$: the time scale needed for dissipation to have an effect is infinite and linear propagation occurs on all spatial scales. For $\eta\neq 0$, on the other hand, we have two regimes, according to the value of the friction coefficient and of the wave number $k$. 
For $k\geq k_0$ we have {\it attenuated} propagating waves, as the frequency has both a real and an imaginary part.
For $k < k_0$ we have {\it evanescent} waves with exponential decay and we recover the overdamped Vicsek-like behavior. 
 
Consistently with the above analysis we see that 
the large scale $k\to 0$ behavior of a system of self-propelled particles is well accounted for by the Vicsek overdamped case and is therefore well described by the hydrodynamic theories of flocking.
There is however a crucial ingredient to be considered when dealing with natural or even artificial groups: flocks are finite. The smallest value of $k$ in the system is $k_\mathrm{min} \sim 1/L$, where $L$ is the linear size of the flock. Hence, even if dissipation is present, if it is sufficiently small it does not affect the scales relevant for information to travel through the group. More precisely, if,
\begin{equation}
\eta < \frac{\sqrt{Jn_c\chi}a}{L} \quad :
\quad
\mathrm{underdamped}\ \mathrm{limit}
\ ,
\label{smammal}
\end{equation}
then there is linear propagation of the information throughout the whole flock. Indeed, for these values of $\eta$ the time scale of the exponential decay is $ \tau >  \sqrt{\chi/(Jn_c a^2)} \ L = L/c_s$. Therefore, small dissipation implies that the damping time constant is larger than the time the information takes to travel across the  flock. In other words, the signal is effectively very weakly damped across the length scale of interest. We conclude that even when a small dissipation is present, propagation of information is qualitatively the same as that described by the zero dissipation theory.

Natural flocks of birds appear to be precisely in this underdamped, quasi-deterministic regime \cite{attanasi+al_14}. Experiments  on large flocks of starlings performing collective turns show that, once the turn (i.e. a direction change) is initiated by some individuals at the edge of the flock, the turning front propagates through the group with a linear dispersion law with very limited attenuation. The relationship (\ref{eq:speed1}) between the experimentally measured speed of information propagation $c_s$ and the alignment strength $J$, predicted by the model in the deterministic limit, is also remarkably verified \cite{attanasi+al_14}. Thus, far from being an idealized construction, the deterministic model introduced previously truly captures some important mechanisms occurring in real biological groups.

\subsection{Transition to the hydrodynamic regime}
\label{sec:transition}
The natural flocks investigated in \cite{attanasi+al_14} live in the quasi-deterministic regime $k>k_0$. Other systems, however, (or even very large flocks) might well explore larger length-scales. This is certainly the case for a variety of fluids or large assemblies of active individuals. As mentioned in the previous section, hydrodynamic theories of flocking offer a sophisticated and detailed description  of the $k\to 0$ regime, which has been investigated in numerical simulations and experiments on granular active matter. While we do not want to give a complete account of this description (we refer the reader to \cite{toner+tu_98}) we briefly comment on what happens as these large length (and time) scales are approached. On the one hand, as we noted in the previous section, when $k\to 0$ the frequency Eq.~(\ref{eq:omega}) becomes purely imaginary indicating that there are not any more purely directional propagating modes. The physical reason is that for times scales $t\gg \tau=2\chi/\eta$  the global spin relaxes due to dissipation and does not act as a quasi-conserved variable, thus the conservation law that determined the propagating nature of the modes is lost. In other terms, the second order terms of the microscopic dynamics become irrelevant, as well as the coupling between direction and curvature. On the other hand, on such large time scales the movement of individuals becomes relevant, and crucial: orientational fluctuations (which would be evanescent  on a fixed network) couple with fluctuations in density. As a consequence, there are propagating  hybridized longitudinal  sound modes transporting both directional and density disturbances. The speed of propagation of such modes usually depends on the average density (as in standard sound), contrary to the purely directional propagating modes of the underdamped regime, where the propagation speed $c_s$ does not depend of the flock density and is  determined by the global alignment strength $J$ (i.e.  the degree of polar order). 
Finally, if the interaction between the moving particles and the surrounding medium becomes important the hydrodynamic theory also predicts shear-bend modes  \cite{toner+tu_98,simha_02} that propagate directional changes. These modes are transverse (i.e. they do not exist in $d=2$) and have limited speed in a co-moving reference frame, and are therefore well distinct from the turning modes observed in flocks, which are longitudinal and have very fast propagation speeds \cite{attanasi+al_14}.

Summarizing, in the quasi-deterministic regime $k>k_0$, where the spin is a quasi-conserved variable, there are propagating modes thanks to the coupling between velocity and spin and the consequent conservation law; these turning modes transport changes of direction {\it and} curvature and describe collective turns in flocks. In the hydrodynamic regime, due to the movement of the network, changes in direction couple with changes in local density generating propagating sound modes; this regime is universal, i.e. independent of the details of the microscopic dynamics, and does not describe transport of curvature.


\section{Numerical simulations}
\label{sec:simulations}

In the previous sections we have used analytical arguments to derive the predictions of the inertial spin model on how orientational information propagates in the ordered phase. We did that, however, working under two assumptions: we disregarded effects due to the reshuffling of the interaction network, and we considered the continuous space limit. While these assumptions are reasonable in certain cases,  we want to test the behavior of the model in full generality. We performed numerical simulations  for a wide range of the relevant parameters $\chi$, $\eta$, and $J$ and explored how the values of these parameters determine the collective dynamics of the system.

We implemented a time discretized version of  the inertial spin model,
\bea
\vec{v}_i(t+dt)&=& \vec{v}_i(t)+ \frac{1}{\chi} \vec{s}_i(t) \times  \vec{v}_i(t) \, dt \\
\vec{s}_i(t+dt) &=& \vec{v}_i(t)+ \vec{v}_i(t) \times \frac{J}{v_0^2} \sum_{j} n_{ij} \vec{v}_j(t) \, dt \\
&-&\frac{\eta}{\chi}s_i(t) \, dt +  \vec{v}_i(t) \times \frac{\vec{\xi}_i(t)}{v_0} \, \sqrt{dt}
\nonumber \\
\vec{r}_i(t+dt)&=& \vec{r}_i(t)+\vec{v}_i(t) \, dt
\label{eq:num}
\eea
where we exploited the equation for $\vec{v}_i$ in order to replace $d\vec{v}_i/dt$ with $\vec{s}_i$, and where we recall $\vec{v}_i\cdot\vec{s}_i=0$. The noise variance is
\begin{equation}
\langle \vec{\xi}_i(t)\cdot \vec{\xi}_j(t')\rangle = (2 d)\,  \eta\,   T \, \delta_{ij} \delta_{t,t'} \ . 
\end{equation}

 \begin{figure*}[t!] 
\centering
 \includegraphics[width=\linewidth]{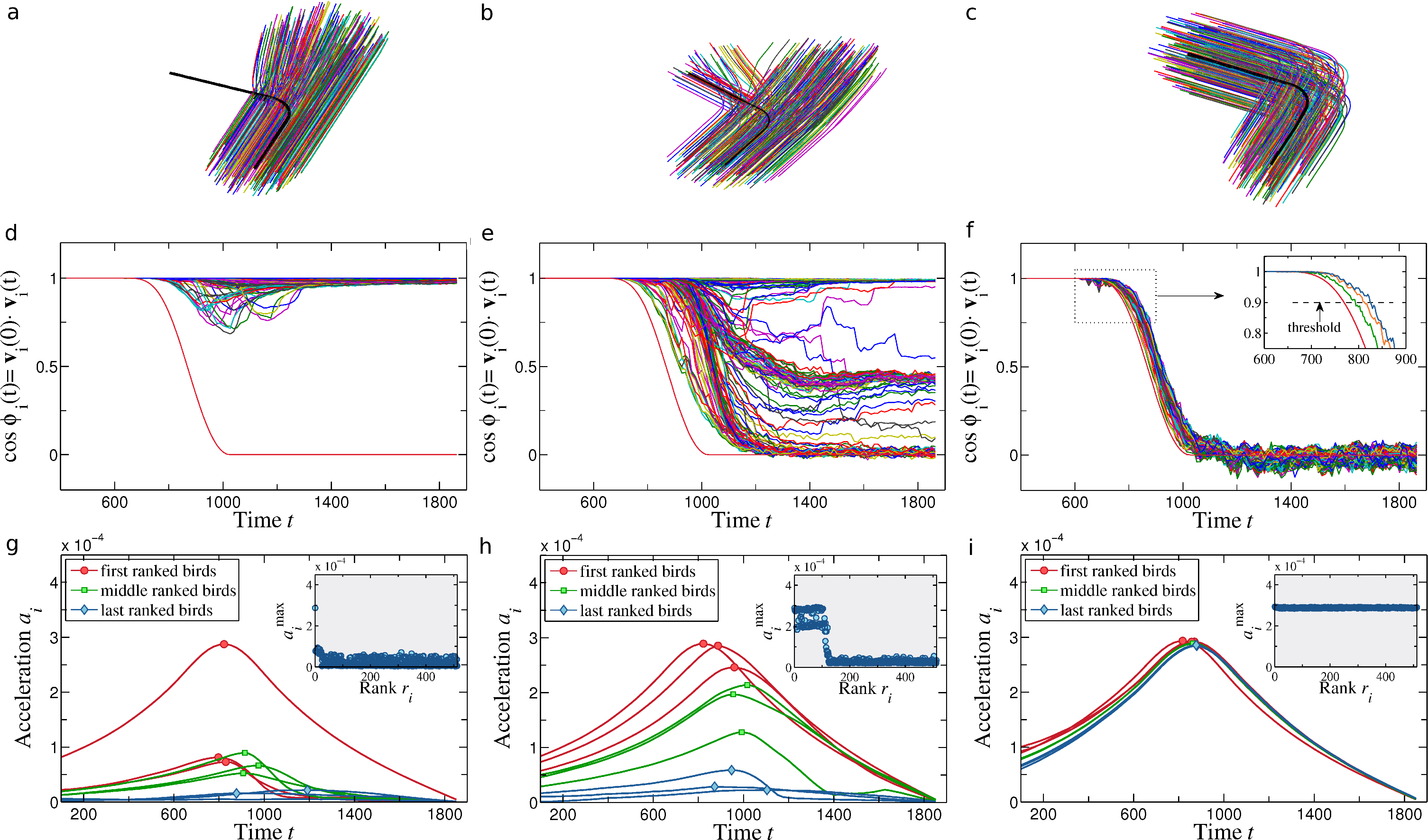}
\caption{Information propagation in different regimes.
{\bf a.} Strongly overdamped regime, $\eta^2/\chi=28.8 \times10^2$. The signal is quickly dissipated and there is no collective turn. All the particles keep following the initial direction of motion without following the initiator. The trajectory of the initiator is displayed as a thick black line.  $\eta=60$, $\chi=1.25$, $J=0.8$. 
{\bf b.} Overdamped regime, $\eta^2/\chi=1.8 \times10^2$. Some propagation occurs, but the signal is strongly attenuated before reaching throughout the whole group. As a consequence, the flock looses cohesion and coherence while turning. $\eta=15$, $\chi=1.25$, $J=0.8$.
{\bf c.} Underdamped regime, $\eta^2/\chi=7.2 \times10^{-2}$. The signal is propagated unattenuated through the whole flock, which performs a neat turn retaining shape and cohesion. $\eta=0.3$, $\chi=1.25$, $J=0.8$. {\bf d}, {\bf e}, {\bf f} Cosine of the individual velocities with the original flight direction of the flock for the three cases displayed in panels a,b,c. The cosine curve of the initiator is displayed as a red curve: at time $t=t_0$ the flight direction of the initiator is tuned from its original direction (coherent with the flock motion) to a final direction following a ramp-like pattern. The inset in panel {\bf f} shows the threshold used to compute the ranking of the particles  (see text).
{\bf g}, {\bf h}, {\bf i}  Individual acceleration profiles for the three cases displayed in panels a,b,c. The acceleration curves have been smoothed with a low-pass filter to cure noise and high frequency oscillations. The insets display the intensity of the peak as a function of time. Particles are ranked according to the order of reaction to the initiator (see text). The other parameters of the simulations are $N=512$, $T=8\times 10^{-5}$, $n_c=6$, $v_0=0.1$. The integration time is chosen as $dt= 0.1 \sqrt{J/\chi}$ to ensure proper simulation time for all values of $J$ and $\chi$.
  }
\label{fig:regimes}
\end{figure*}

Our objective is twofold: on the one hand we want to check the model's behavior in the ordered phase in the various possible regimes; on the other hand we want to compare it to results obtained in natural flocks of birds \cite{attanasi+al_14}. To this end, we chose the connectivity network between particles  according to a topological rule, as in real flocks \cite{cavagna+al_10,bialek+al_12}: each particle interacts with the first $n_c$ neighbors so that $n_{ij}=1$ if $j$ is among the first $n_c$ neighbors of $i$, and zero otherwise.  We focus on finite groups of particles with open boundary conditions, but qualitatively similar results also hold for periodic boundary conditions.

We are interested in understanding how, for different values of $\eta$, $\chi$ and $J$,  directional information propagates in the system, and whether  the simulated flock is able to sustain coherent collective turns as natural flocks do.  In real flocks collective turns start locally: an individual bird, the initiator, starts to turn and then the information about turning propagates through the system with a linear dispersion law \cite{attanasi+al_14}. To mimic this situation, we consider the system at low enough values of noise as to generate an ordered flock with large polarization ($\Phi > 0.95$ as in natural flocks), and we initialize all the $\vec{s}_i=0$ so that the flock as a whole performs a straight motion. Then, we chose at random a particle inside the flock and make it artificially turn changing its flight direction with a ramp-like time dependence (red line in Fig.~\ref{fig:regimes}d,e,f). The change in direction of the initiator affects - through Eqs.~(\ref{eq:num}) - the dynamics of the spins of nearby particles, which might or might not cause a collective movement. We then look at the behavior of the whole flock in time to assess whether the signal has propagated, and what kind of dispersion law is obeyed.

According to the analytical arguments given in the previous section, we expect two different regimes, according to the values of $\eta$ and $\chi$:
\begin{itemize}
\item[1)] {\bf Overdamped regime:} $\eta^2/\chi > n_c J (a/L)^2 $. In this regime, given a system of size $L$, some attenuated propagation occurs up to certain spatial scales ($k \ge k_0$, with $k_0=1/(2 a \sqrt{n_cJ\chi})>1/L$). On larger scales ($k<k_0$) however, dissipation takes over leading to an exponential decay of the signal. The extreme case occurs for $\eta^2/\chi\to\infty$ (or $\chi/\eta^2\to 0$), corresponding to the Vicsek model, when propagation of orientational perturbations does not occur (unless we reach hydrodynamic lenght-scales where density and orientational modes couple, see \cite{toner+tu_98,toner+al_98} and Sec.~\ref{sec:transition}). 
\item[2)] {\bf Underdamped regime:} $\eta^2/\chi\ll n_c J (a/L)^2$. In this regime there is linear propagation of the signal throughout the whole system with negligible attenuation. The speed of information propagation is determined solely by the ratio of alignment to inertia, $c_s=a \sqrt{n_c J/\chi}$.
\end{itemize}

\begin{figure*}[t!] 
  \centering
  \includegraphics[width=1.8\columnwidth]{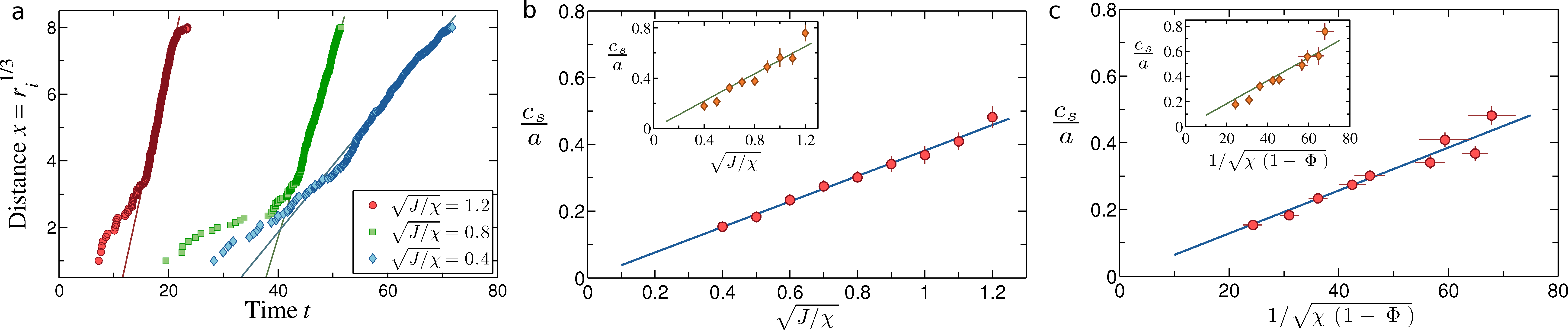} 
 \caption{Propagation curve and dispersion law.
{\bf a.} Propagation curve. Distance $x$ traveled by the turning front vs time, for three different values of the parameters $\eta$, $\chi$, $J$ in the small dissipation regime. The distance $x$ traveled in a time $t$ is proportional to the radius of the sphere containing the first $r(t)$ birds in the rank, namely $x(t)=[r(t)/\rho]^{1/3}$.  $t=0$ corresponds to the time when the initiator starts the turn. The three curves correspond respectively to $\eta=0.3$, $\chi=0.83$, $J=1.2$ (red circles); $\eta=0.3$, $\chi=1.25$, $J=0.8$ (green squares); $\eta=0.3$, $\chi=2.50$, $J=0.4$ (blue diamonds). The rank is computed using the cosine curves (see text). The speed of propagation, $c_s$, is the slope of the propagation curve in linear regime. The colored straight lines show the linear fits for the three different curves. 
{\bf b.} Speed of propagation $c_s$, normalized by the typical distance $a$ of the interacting neighbors, as a function of $\sqrt{J/\chi}$. For each value of $J/\chi$ we run several simulations and estimate $c_s$  from the slope of the propagation curve, the point corresponds to the average value and bars to standard error.  The speed of propagation depends on the ratio $J/\chi$ as predicted by the analytic arguments in the previous section: more ordered flocks transmit the turning information quicker. Inset: same curve as in main panel, but using accelerations curves to compute ranks and propagation curves.  
{\bf b.} Speed of propagation $c_s$, normalized by $a$, as a function of  $1/\sqrt{\chi(1-\Phi)}$. Inset: same curve as in main panel, but using accelerations curves to compute ranks and propagation curves.
 }
\label{fig:rank}
\end{figure*}

We performed simulations for values of the parameters corresponding to these two different regimes. As can be seen from Fig.~\ref{fig:regimes},
what we observe is fully consistent with what is described above. The top panels in this figure display the trajectories of the individual particles in the flock  for three decreasing values of $\eta^2/\chi$,  the trajectory of the initiator being represented by the thick black line. The intermediate panels show, for the same events, the cosine of the velocity of each particle with the original flight direction of the flock. Here the initiator is represented as a red line, and one can see the ramp-like dependence of the perturbation. The lower panels reproduce the individual acceleration profiles, and are particularly useful to determine if and how particles turn, and how the turning signal is attenuated during time. 

The first two cases in Fig.~\ref{fig:regimes} (panels a,b) belong to the overdamped regime. For values of $\eta^2/\chi$ much larger than $n_c J (a/L)^2$ (which is of order 1 for our simulations) the turn of the initiator does not trigger any collective turn, since the signal does not propagate and is quickly damped (panel a). We can see from the cosine curves (panel d) that some of the neighbors feel the disturbance caused by the initiator, but this effect is very small: no particle follows the initiator, no collective turn is triggered and the flock ultimately keeps its original flight direction (see also attached Video 1). For smaller values of $\eta^2/\chi$ (panel b) there is some propagation, but it is progressively attenuated as the signal travels through the group and eventually decays (panels e,h). Some particles are reached by the directional perturbation before the signal is damped (corresponding to scales $k>k_0$), and follow the initiator changing their flight direction and performing a turn. Most of the particles, however, are not significantly affected by the perturbation, consistently with a strong damping of the signal on large enough scales.  This effect is clearly seen by looking at the acceleration profiles of the particles (Fig.~\ref{fig:regimes}h). Some particles display a profile similar to the one of the initiator with a clear peak marking the occurrence of a turn. The height of the peak, however, decreases as the turning information is transmitted (i.e. particles start turning one after the other) and at some point it abruptly drops to zero (see inset).  
In this case, the flock breaks into two subgroups: the largest component keeps the original flight direction, while a smaller fragment is dragged by the initiator (see attached Video 2).  In this respect we note that, due to active nature of the particles, the effect of overdamping is even stronger than discussed in sec.~\ref{sec:propagation}: not only the turning information fails to percolate through the group, but the flock breaks up and looses cohesion. 

In the underdamped regime (panel c), on the contrary, there is robust propagation of the signal through the whole group: all the particles follow the initiator changing their flight direction with negligible attenuation. Both the cosine curves and the acceleration profiles show that the particles turn in a highly coordinated way: the curves are very similar one to the other with a shift in time corresponding to the propagation of the turn from the initiator to the last particle (Fig.~\ref{fig:regimes}f,i). The flock performs a neat collective turn retaining shape and cohesion during motion (Fig.~\ref{fig:regimes}c, and Video 3).

In this regime, we can quantify the dispersion law using the method introduced in \cite{attanasi+al_14}. We rank all the particles according to their order of turning: the initiator has rank $r=0$, then the particle that first starts turning after the initiator has rank $r=1$ and so on. To determine the rank of a particle we follow two different procedures. We look at the cosine curve of the particle and determine its absolute turning delay from the initiator (and therefore its rank) as the time when the cosine reaches a threshold value (we choose 0.9 - see inset in Fig.~\ref{fig:regimes}f).   Alternatively, we proceed as in  \cite{attanasi+al_14}: from the acceleration curves we compute the relative time delays between pairs of particles as the shift necessary to superimpose their curves. Once the relative delays are found for all particles we use a standard algorithm to rank them all. This method is more robust when it is not known a priori who is the initiator (as in experiments \cite{attanasi+al_14}). The two procedures give equivalent results.

If we now look at the rank as a function of delay time we get the ranking curve, describing how the turn is executed through the group. Since the turn starts locally and the flock moves in three dimensions, $r^{1/3}(t)$ is a measure of the distance $x$ travelled by the turning wave in time $t$. As can be seen from Fig.~\ref{fig:rank}a, the distance vs. time curve displays a clear linear dependence corresponding to a linear dispersion law, as predicted by the analytic argument in previous section. We can also check quantitatively the prediction for the propagation speed
\be
c_s=a \, \sqrt{\frac{n_c J}{\chi}}
\label{eq:speed}
\ee 
To do so, we run numerical simulations for several values of the parameters in the underdamped regime. We fix $\eta=0.3$ and vary $J$ and $\chi$ so as to obtain propagation curves with different slopes, but keep $J \chi$ constant in order to have the same value of $k_0<1/L$. 
Then, for each simulation, we compute the distance vs. time curve as in Fig. ~\ref{fig:rank}a and we estimate $c_s$ from a fit of the linear part of the curve. As can be seen from Fig.~\ref{fig:rank}b, the dependence of $c_s$ on $J/\chi$ is very well reproduced by numerical data. 

An alternative way to check the speed dependance - which is particularly useful when comparing to real data - is to plot it as a function of the polarization. We know from Section \ref{pussy} that the statistical properties of the velocities are described through the Boltzmann measure Eq.~(\ref{eq:boltzmann}). If we use this distribution, we can compute the expected value of the polarization in the ordered phase. This gives $\Phi=1-T\,J^{-1} \mathrm{Tr}\Lambda^{-1}$ (see \cite{bialek+al_12} for the details of the computation). We can therefore eliminate $J$ in favour of $\Phi$, and obtain,
\be
c_s=a\,  \sqrt{\frac{n_c\, T\; (\mathrm{Tr}\Lambda^{-1})}{(1-\Phi) \chi}}  \ .
\label{eq:speedu}
\ee 
The parameters $T$ and $n_c$ are kept constant in our simulations; moreover, the trace of the inverse discrete Laplacian $\Lambda_{ij}$ fluctuates little from sample to sample. Therefore, to check equation (\ref{eq:speedu}) we plot the speed propagation $c_s/a$ as a function of $1/\sqrt{(1-\Phi)\chi}$ (see Fig.~\ref{fig:rank}c). Also in this case the numerical data agree very well with the analytical prediction. 

The behavior displayed in Fig.~\ref{fig:rank} is analogous to what is observed in natural flocks of birds. The propagation curves look indeed very similar to the ones computed for real flocks and shown in \cite{attanasi+al_14}. Even more importantly, Fig.~\ref{fig:rank}c is remarkably similar to what found in real data, and the model correctly reproduces the experimentally found relationship between propagation speed and degree of order in the system. We can therefore conclude that Eqs. (\ref{eq:model}) in the underdamped limit fully describe the correlated collective turning exhibited by real flocks. We remind that in this limit inertia and deterministic effects dominate over dissipation: it is the Hamiltonian structure of the dynamical equations, and the connected conservation law, that cause the linear dispersion law and the highly non-trivial relationship between speed of propagation and degree of order in the system.


\section{Conclusions}

The main point of our work is that momentum matters. More precisely, the powerful mathematical entanglement between symmetry, momentum and conservation, is the crucial ingredient giving rise to a dispersion relation, namely to a propagation law, identical to the one observed in real flocks of birds during global changes of direction. If momentum is disregarded, the dispersion relation changes dramatically and linear propagation collapses: the practical result is that no collective change of direction can be achieved by the group. 

But what momentum is that? We hope we have convinced the reader that the momentum essential to our new model is {\it not} linear momentum, {\it nor} orbital angular momentum. Accordingly, the generalized inertia in our equation is not the mass, nor the moment of inertia. Linear momentum conservation is clearly violated in active matter systems: this is because the particles are, indeed, active. Momentum (and energy) is injected and dissipated continuously at the individual level. For this reason describing active matter systems in terms of Hamiltonian dynamics in the $(x,p)$ variables is impossible. But once we wrap all the complicated active mechanism into an effective constraint for the velocity, we can define an effective angular variable, the phase, which parametrizes the rotation of the velocity and satisfies the constraint. The relevant momentum then is the generator of this internal rotation, that is the spin. The resistance of the particle to changes in its spin is the turning inertia. This set of variables, $(\varphi, s, \chi)$, has nothing to do with the more familiar set, $(\theta, l, I)$, of polar angle, orbital angular momentum and moment of inertia.

This unusual definition of momentum is the precise reason why the propagation phenomenon described by the inertial spin model is so akin to superfluidity.
One could object that linear propagation of information is in fact the most mundane thing of all, sound waves being its most obvious example.
The point is that sound waves transport density fluctuations, the symmetry involved is translation, and the relevant momentum is $p$.
On the other hand, second sound waves, those associated to superfluidity, transport phase fluctuations, the symmetry is internal rotation, and momentum is $s$, the spin. This transport of phase and spin {\it is} typical of superfluid systems that transport quantum phase. However weird this connection between flocks and quantum systems may seem, it is mathematically exact.

What is surprising is that the new phenomenology of linear propagation is obtained at the rather low price of introducing just one more parameter to the standard Vicsek model, that is the spin inertia $\chi$. Moreover, because of the standard Hamiltonian decoupling between coordinates and momenta, the static (short time)
properties of the inertial spin model are exactly the same as the Vicsek model, so that we do not have to worry about how to recover the rich physical description of static correlations previously achieved. In the overdamped limit momentum and inertia become irrelevant and ones recovers the original Vicsek dynamics. Finally, the large length-scale behavior is described by the standard hydrodynamics of flocking \cite{toner+tu_98}. 
Hence, the inertial spin model introduced here achieves quite a lot in terms of new and experimentally accurate phenomenoogy, at a very little cost in terms of modeling complexity. 

At the more concrete numerical level, we have clearly shown that the inertial spin model describes turns accurately. A change of direction can successfully be initiated by just {\it one} bird and the turn propagates very efficiently to the rest of the flock. The speed of propagation is related to the strength of the alignment interaction and to the turning inertia exactly in the way predicted by our mathematical equations and observed in real experiments \cite{attanasi+al_14}. On the contrary, if spin conservation is neglected (overdamped or Vicsek case) no collective turn is achieved. Interestingly, due to the active nature of flocking, the effect of overdamping is even more dramatic than predicted in the fixed-lattice case: instead of a diffusive and damped propagation of the signal, one observes no propagation at all, as the failure to transport the turning information ultimately produces the break up of the flock.

An interesting question is what happens when considering very large systems and time scales. This is the realm addressed by the hydrodynamic theories of flocking \cite{toner+tu_95,toner+tu_98,ramaswamy_review}. On these scales the motion of the network cannot be disregarded and a non-trivial hybridization occurs between density modes (caused by the relative movement of the particles) and orientational modes (due to alignment between velocities).  As we already discussed in the paper, if some rotational dissipation is present in the system (i.e. $\eta\ne0$), the asymptotic long wavelength behavior of our model is the same as the overdamped Vicsek limit, and we therefore expect the predictions of current hydrodynamic theories to describe it appropriately. One can however wonder whether a hydrodynamic description exists that  also accounts for the underdamped behavior occurring at smaller scales, where many relevant phenomena like collective turns in flocks take place.  To do so, one would need to modify the hydrodynamic equations to include a weakly damped spin field. This opens new  interesting perspectives at the theoretical level.

What we have done here and in Reference \cite{attanasi+al_14} is to push one step further the ambitious program to tame the vast richness of biological phenomena using the powerful conceptual framework of theoretical physics. This program is of course not new. Kinetic approaches \cite{bertin+al_06, ihle_11} and hydrodynamic theories of active matter \cite{toner+tu_98,ramaswamy_review,marchetti_review} used tools as kinetic theory, Navier-Stokes equations and dynamical renormalization group to determine the long distance and large time scaling properties of flocks and to explain why the Mermin-Wagner theorem is violated. 
At the static level, the maximum entropy approach \cite{bialek+al_12,bialek+al_13,cavagna+al_13} endeavors to define a thermodynamic description of biological collective systems based exclusively on the experimental data. In this approach, theoretical physics tools, as the Goldstone theorem and scaling relations, have powerful consequences \cite{cavagna+al_10}.

Here, however, we do something different. We bring into the arena the abstract power of symmetries and conservation laws in linking static properties to dynamical predictions at the level of the microscopic description of the system. Symmetry is not necessarily a dynamical concept: it is a very general and profound condition that a system may or may not enjoy. However, the conservation laws predicted by symmetry through Noether's theorem and Hamilton's description have deep dynamical implications, which, as we have seen, can be experimentally verified in natural system and in numerical simulations alike. 
To do this the right set of {\it effective} variables must be found - a recurrent and quite natural theme in physics. We believe that this kind of approach may have some generality in biological systems other than flocks.

\vskip 0.3 truecm
{\bf Acknowledgments.}
We thank William Bialek, Serena Bradde, Paul Chaikin and Dov Levine for discussions. Work in Rome was supported by grants IIT--Seed Artswarm, ERC--StG n.257126 and US-AFOSR - FA95501010250 (through the University of Maryland). Work in Paris was supported by grant ERC--StG n.
306312.
\vskip 0.3 truecm




\begin{thebibliography}{99} 


\bibitem{camazine+al_01}
Camazine, S., Deneubourg, J.-L., Franks, N.~R., Sneyd, J., Theraulaz, G., and
  Bonabeau, E.  \textit{{Self-Organization} in Biological Systems}
  (Princeton University Press, Princeton, 2001)

\bibitem{couzin+krause_03} Self-organization and collective behavior in vertebrates
Couzin, I.~D. Krause, J.    \textit{Adv  Study  Behavi} {\bf 32}: 1--75 (2003.)

\bibitem{giardina_08} 
Giardina I.  Collective behavior in animal groups: theoretical models and empirical studies {\em HFSP Journal} {\bf 2}: 205--219 (2008)

\bibitem{sumpter_10} Sumpter DJT. Collective Animal Behavior (Princeton University Press, Princeton 2010)

\bibitem{cavagna+giardina_14} Cavagna A, Giardina I, Bird Flocks as Condensed Matter, {\it Ann. Rev. Cond. Matt. Phys.} DOI: 10.1146/annurev-conmatphys-031113-133834 (2014).



\bibitem{aoki_82} Aoki I., A
simulation study on the schooling mechanism in fish.
{\it Bull. Jpn. Soc. Sci. Fish.} {\bf 48}: 1081-1088 (1982)


\bibitem{reynolds_87} Reynolds CW. 1987. {\it Computer Graphics} {\bf 21}: 25-33.

\bibitem{huth_92} Huth, A., \& Wissel, C.  The Simulation of the
Movement of Fish Schools.
{\em J. Theor.  Biol. } {\bf 156}, 365--385 (1992).

\bibitem{couzin+al_02}
Couzin, I.~D., Krause, J., James, R., Ruxton, G.~D., and Franks, N.~R.
  Collective memory and spatial sorting in animal groups \textit{J.
Theor. Biol.}
   \textbf{218}, 1--11 (2002)






\bibitem{hemelrijk_10}
Hildenbrandt, H., Carere, C., and Hemelrijk, C. Self-organized aerial displays of thousands of starlings: a model \textit{Behav. Ecol.}   \textbf{21}: 1349--1359. (2010) 






\bibitem{vicsek+al_95}
Vicsek, T., Czir{\'o}k, A., Ben-Jacob, E., Cohen, I., and Shochet, O.
  Novel type of phase transition in a system of self-driven particles
  \textit{Phys. Rev. Lett.} \textbf{75}, 1226--1229 (1995).

\bibitem{toner+tu_95}Toner, J. and Tu, Y.  Long-range order in a two-dimensional dynamical XY model: how birds fly together
 \textit{Phys. Rev. Lett.}  \textbf{75}: 4326--4329 (1995)

\bibitem{gregoire+chate+tu_03} G  Gr\'egoire, H Chat\'e \& Y Tu,  Moving and staying together without a leader {Physica D} {\bf 181}: 157-170 (2003)

\bibitem{gregoire+chate_04}
Gr{\'e}goire, G. and Chat{\'e}, H. Onset of collective and cohesive motion
  \textit{Phys. Rev. Lett.} \textbf{92}, 025702 (2004)



\bibitem{dorsogna+al_06}
DÕOrsogna MR, Chuang YL, Bertozzi AL, and Chayes LS.  Self-propelled particles with soft-core interactions: patterns, stability, and collapse{\it Phys Rev Lett}{\bf  96}, 104302. (2006)


\bibitem{huepe+al_07} 
Aldana A, Dossetti V., Huepe C, Kenkre VM, Larralde H. . Phase transitions in systems of self-propelled agents and related network models {\it Phys. Rev. Lett.} {\bf 98}: 095702 (2007)


\bibitem{ginelli+chate_10} Ginelli, F. \& Chat\'e, H.  Relevance of metric-free interactions in flocking phenomena. {Phys. Rev. Lett.} {\bf 105}: 168103 (2010)




\bibitem{toner+tu_98}
Toner, J. and Tu, Y. Flocks, herds, and schools: A quantitative theory of
  flocking \textit{Phys. Rev. E} \textbf{58}, 4828--4858 (1998)


\bibitem{vicsek_review}  Vicsek T, Zafeiris A. Collective motion {\it Physics Reports}  {\bf 517}: 71Ð140 (2012)


\bibitem{ramaswamy_review} Ramaswamy S.  The Mechanics and Statistics of Active Matter  {\it Ann. Rev. Cond. Matt. Phys.} 1, 301 (2010).

\bibitem{marchetti_review} Marchetti MC, Joanny JF, Ramaswamy S, Liverpool TP, Prost J et al. Hydrodynamics of soft active matter {\it Rev. Mod. Phys.} {\bf 85} 1143  (2013). 



\bibitem{toner+al_98}Tu Y., Toner J., Ulm M.  Sound waves and the absence of galilean invariance in flocks. {\it Phys. Rev. Lett.}  {\bf 80} 4819 (1998).

Toner, J. and Tu, Y.  Long-range order in a two-dimensional dynamical XY model: how birds fly together
 \textit{Phys. Rev. Lett.}  \textbf{75}: 4326--4329 (1995)



\bibitem{krishna_04} Justh, E.W., \& Krishnaprasad, P.S. {\em
Equilibria and steering laws for planar formations},
Systems \& Controls Letters {\bf 52}, 25--38 (2004)



\bibitem{tanner+al_07} Tanner, H.G.; Jadbabaie, A.; Pappas, G.J., Flocking in Fixed and Switching Networks, {\it Automatic Control, IEEE Transactions} {\bf 52},  863--868 (2007)


\bibitem{attanasi+al_14} Attanasi A, Cavagna A, Del Castello L, Giardina I, Jelic A, Melillo S, Parisi L, Shen E, Viale M. Information transfer and behavioural inertia in starling flocks {\it Nat. Phys.} {\bf 10}, 691--696 (2014) 



\bibitem{bialek+al_12}
Bialek, W., Cavagna, A., Giardina, I., Mora, T., Silvestri, E., Viale, M., and
  Walczak, A.~M.  Statistical mechanics for natural flocks of birds
  \textit{Proc. Natl. Acad. Sci. USA} \textbf{109},
  4786--4791 (2012)

\bibitem{cavagna+al_10}
Cavagna, A., Cimarelli, A., Giardina, I., Parisi, G., Santagati, R., Stefanini,
  F., and Viale, M. Scale-free correlations in starling flocks.  \textit{Proc. Natl. Acad. Sci. USA} \textbf{107}:11865--11870 (2010)


\bibitem{ballerini+al_08a}
Ballerini, M., Cabibbo, N., Candelier, R., Cavagna, A., Cisbani, E., Giardina,
  I., Lecomte, V., Orlandi, A., Parisi, G., Procaccini, A., et~al.
  Interaction ruling animal collective behavior depends on topological rather
  than metric distance: Evidence from a field study
  \textit{Proc. Natl. Acad. Sci. USA} \textbf{105}, 1232--1237 (2008).



\bibitem{goldstein_80} Goldstein H., Classical Mechanics (Addison-Wesley Publishing Company, Reading, MA, 1980)

\bibitem{fetter+walecka_12} Fetter AL, Walecka, JD. Theoretical Mechanics of Particles and Continua ( Courier Dover Publications, 2012)




\bibitem{heppner_92} Pomeroy, H, Heppner F.  Structure of turning in airborne rock dove (Columba livia) flocks
 {\em Auk} {\bf 109}: 256Ð267 (1992)
 


\bibitem{duarte+al_13} Cavagna A., Duarte Queir—s SM,  Giardina I, Stefanini F and Viale M.  Diffusion of individual birds in starling flocks {\it Proc. R. Soc. B} {\bf 280}, 20122484 (2013)






\bibitem{halperin+hohenberg_77} Hohenberg, P. C. and Halperin, B. I., Theory of dynamic critical phenomena,
 {\it Rev. Mod. Phys.} {\bf 49}, 435--479 (1977)
 
\bibitem{halperin_69} Halperin, B.I.  \& Hohenberg, P.C. Hydrodynamic
Theory of Spin Waves, {\em Phys. Rev.} {\bf 188}, 898--918 (1969)

\bibitem{sonin_10} Sonin, E.B. Spin currents and spin superfluidity,
{\em Advances in Physics} {\bf 59}, 181--255 (2010)


\bibitem{matsubara+matsuda_56b} Matsubara, T. \& Matsuda, H., A lattice model of Liquid Helium, I,  {\it Prog. Theor. Phys.} {\bf 16}, 569--582 (1956)

\bibitem{helium_47} Lane, C.T., Fairbank, H.A. \& Fairbank, W.M.
Second Sound in Liquid Helium II, {\em Phys. Rev.} {\bf 71}, 600--605
(1947)

\bibitem{krishnaprasad_04}
Justh, E.W. \& Krishnaprasad, P.S. Equilibria and steering laws for planar formations, {\it Systems \& Controls Letters} {\bf 52}, 25--38 (2004).

\bibitem{szabo_09}
Szabo, P., Nagy, M., \& Vicsek, T.
Transitions in a self-propelled-particles model with coupling of accelerations
{\it  Phys. Rev. E} {\bf 79}, 021908 (2009)

\bibitem{hemelrijk_11}
Hemelrijk C.K. \& Hildenbrandt H. 
Some causes of the variable shape of flocks of
  birds \textit{PLoS ONE} \textbf{6}, e22479 (2011)

\bibitem{gautrais+al_12}
Gautrais, J., Ginelli, F., Fournier, R., Blanco, S., Soria, M.,
Chate\'e, H. \& Theraulaz, G. Deciphering interactions in moving
animal groups, {\it  Plos Comp. Biol.} {\bf 8} e1002678 (2012)


\bibitem{sumino+al_12} Sumino Y., Nagai K.H., Shitaka Y., Tanaka D., Yoshikawa K., Chat\'e H., Oiwa K. Large-scale vortex lattice emerging from collectively moving microtubules. {\it Nature} {\bf 483}, 448--452 (2012)


\bibitem{zwanzig} Zwanzig, R. Nonequilibrium statistical mechanics (Oxford University Press, Oxford, 2001).

\bibitem{gardiner} Gardiner, C. W. Handbook of stochastic methods. Vol. 3.(Springer, Berlin, 1985).

\bibitem{goldstone}Goldstone, J. Field theories with Superconductor solutions. {\it Il Nuovo Cimento} {\bf 19}, 154--164 (1961)


\bibitem{simha_02} Ramaswamy S.,  Simha R.A., Hydrodynamics Fluctuations and Instabilities in Ordered Suspensions of Self-
Propelled Particles, {\it Phys. Rev. Lett.}{\bf 89}, 058101 (2002)


\bibitem{camperi+al_12} Camperi M, Cavagna A, Giardina I, Parisi G, Silvestri E,   Spatially balanced topological interaction grants optimal cohesion in flocking models {\it Interface Focus}{\bf 2}, 715--725 (2012)

\bibitem{ref-spinwave} FJ Dyson, General theory of spin--wave interactions. {\em Phys Rev } {\bf 102}, 1217--1230 (1956).


\bibitem{bertin+al_06}
Bertin E, Droz M, and Gregoire G.  {\it Phys. Rev. E} {\bf 74}, 022101 (2006)

\bibitem{ihle_11} Ihle T, Kinetic theory of flocking: Derivation of hydrodynamic equations. {\it  Phys. Rev. E}{\bf 83} 030901 (2011)


\bibitem{bialek+al_13}
Bialek, W., Cavagna, A., Giardina, I., Mora, T.,  Pohl O., Silvestri, E., Viale, M., and Walczak, A.~M. 
Social interactions dominate speed control in poising natural flocks near criticality.  {\it Proc. Natl.  Acad.  Sci. USA} {\bf 111}, 7212-7217 (2014)


\bibitem{cavagna+al_13}
Cavagna, A., Giardina, I., Ginelli I, Mora, T., Piovani D., Tavarone R.  and Walczak, A.~M.  
Dynamical maximum entropy approach to flocking.  {\it Phys. Rev. E} {\bf 89} 042707  (2014)





\end{thebibliography}
\end{document}